\input harvmac

\input epsf
\ifx\epsfbox\UnDeFiNeD\message{(NO epsf.tex, FIGURES WILL BE IGNORED)}
\def\figin#1{\vskip2in}
\else\message{(FIGURES WILL BE INCLUDED)}\def\figin#1{#1}\fi
\def\ifig#1#2#3{\xdef#1{fig.~\the\figno}
\goodbreak\midinsert\figin{\centerline{#3}}%
\smallskip\centerline{\vbox{\baselineskip12pt
\advance\hsize by -1truein\noindent\footnotefont{\bf
Fig.~\the\figno:} #2}}
\bigskip\endinsert\global\advance\figno by1}

\def\footnotefont{\tenpoint}

\newwrite\ffile\global\newcount\figno \global\figno=1
\def\fig{fig.~\the\figno\nfig}
\def\nfig#1{\xdef#1{fig.~\the\figno}%
\writedef{#1\leftbracket fig.\noexpand~\the\figno}%
\ifnum\figno=1\immediate\openout\ffile=figs.tmp\fi\chardef\wfile=\ffile%
\immediate\write\ffile{\noexpand\medskip\noexpand\item{Fig.\ \the\figno. }
\reflabeL{#1\hskip.55in}\pctsign}\global\advance\figno
by1\findarg}

\overfullrule=0pt
\parindent 25pt
\tolerance=10000

\def\RR{{$R\otimes R$}}
\def\ZZ {{\bf Z}}

\def\G(#1){\Gamma(#1)}

\def\threeh{{\scriptstyle {3 \over 2}}}
\def\fiveh{{\scriptstyle {5 \over 2}}}
\def\sevenh{{\scriptstyle {7 \over 2}}}

\def\half{{\scriptstyle {1 \over 2}}}

\def\(#1#2){(\zeta_#1\cdot  \zeta_#2)}
\def\ls{l_s}

\def\hn{{\hat n}}
\def\hm{{\hat m}}

\def\hlambda{\hat \lambda}
\def\hrho{\hat \rho}
\def\hsigma{\hat \sigma}
\def\hdelta{\hat \Delta}

\def\hDelta{\hat \Delta}
\def\hK{\hat K}

\def\C|#1{{\cal #1}}
\def\calV{{\cal V}}
\def\calR{{\cal R}}
\def\calT{{\cal T}}
\def\GST{{\cal G}_{ST}}
\def\GTU{{\cal G}_{TU}}
\def\GUS{{\cal G}_{US}}

\def\calgst{\GST^s}
\def\calgtu{\GTU^s}
\def\calgus{\GUS^s}
\def\calw{{\cal W}^s}

\def\calK{{\cal K}}
\def\calF{{\cal F}}
\def\calZ{{\cal Z}}

\def\calE{{\cal E}}

\lref\Apostol{
Tom M. Apostol, {\sl Introduction to Analytic Number Theory}, (1976) Springer-Verlag, New York
}

\lref\Terras{A. Terras, {\sl Harmonic analysis on symmetric spaces and applications I \& II}, Springer-
Verlag 1985.
}
\lref\IengoPR{
  R.~Iengo,
  {\sl Computing the R**4 term at two super-string loops,}
  JHEP {\bf 0202}, 035 (2002)
  [arXiv:hep-th/0202058].
}

\lref\SinhaZR{
  A.~Sinha,
  {\sl The $\hat G^4\lambda^{16}$ term in IIB supergravity,}
  JHEP {\bf 0208}, 017 (2002)
  [arXiv:hep-th/0207070].
}

\lref\DamourZB{
  T.~Damour and H.~Nicolai,
  {\sl Higher order M theory corrections and the Kac-Moody algebra E(10),}
  Class.\ Quant.\ Grav.\  {\bf 22}, 2849 (2005)
  [arXiv:hep-th/0504153].
}
\lref\rfMatone{
M. Matone and R. Volpato, {\sl Higher Genus Superstring Amplitudes from the
Geometry of Moduli Space}, arXiv:hep-th/0506231.
}
\lref\zhou{
  Z.~G.~Xiao and C.~J.~Zhu,
  {\sl Factorization and unitarity in superstring theory,}
  JHEP {\bf 0508}, 058 (2005)
  [arXiv:hep-th/0503248].\hfill\break
  W.~J.~Bao and C.~J.~Zhu,
  {\sl Comments on two-loop four-particle amplitude in superstring theory,}
  JHEP {\bf 0305}, 056 (2003)
  [arXiv:hep-th/0303152].\hfill\break
   Z.~J.~Zheng, J.~B.~Wu and C.~J.~Zhu,
  {\sl Two-loop superstrings in hyperelliptic language. III: The four-particle amplitude,}
  Nucl.\ Phys.\ B {\bf 663}, 95 (2003)
  [arXiv:hep-th/0212219].\hfill\break
  Z.~J.~Zheng, J.~B.~Wu and C.~J.~Zhu,
  {\sl Two-loop superstrings in hyperelliptic language. II: The vanishing of the
  cosmological constant and the non-renormalization theorem,}
  Nucl.\ Phys.\ B {\bf 663}, 79 (2003)
  [arXiv:hep-th/0212198].\hfill\break
  Z.~J.~Zheng, J.~B.~Wu and C.~J.~Zhu,
  {\sl Two-loop superstrings in hyperelliptic language. I: The main results,}
  Phys.\ Lett.\ B {\bf 559}, 89 (2003)
  [arXiv:hep-th/0212191].
}

\lref\dgp{
  E.~D'Hoker, M.~Gutperle and D.~H.~Phong,
 {\sl Two-loop superstrings and S-duality,}
  arXiv:hep-th/0503180.
}
\lref\dhokerphonglecture{
  E.~D'Hoker and D.~H.~Phong,
  {\sl Lectures on two-loop superstrings,}
  arXiv:hep-th/0211111.
}

\lref\dhokerphong{
  E.~D'Hoker and D.~H.~Phong,
  {\sl Two-loop superstrings. VI: Non-renormalization theorems and the 4-point
  function,}
  Nucl.\ Phys.\ B {\bf 715}, 3 (2005)
  [arXiv:hep-th/0501197];\hfill\break
{\sl Two-loop superstrings. V: Gauge slice independence of the N-point
  function,}
  Nucl.\ Phys.\ B {\bf 715}, 91 (2005)
  [arXiv:hep-th/0501196];\hfill\break
{\sl Two-loop superstrings. IV: The cosmological constant and modular forms,}
  Nucl.\ Phys.\ B {\bf 639}, 129 (2002)
  [arXiv:hep-th/0111040];\hfill\break
  {\sl Two-loop superstrings. III: Slice independence and absence of
  ambiguities,}
  Nucl.\ Phys.\ B {\bf 636}, 61 (2002)
  [arXiv:hep-th/0111016];\hfill\break
{\sl Two-loop superstrings. II: The chiral measure on moduli space,}
  Nucl.\ Phys.\ B {\bf 636}, 3 (2002)
  [arXiv:hep-th/0110283];\hfill\break
  {\sl Two-loop superstrings. I: Main formulas,}
  Phys.\ Lett.\ B {\bf 529}, 241 (2002)
  [arXiv:hep-th/0110247].
}

\lref\BerkovitsMultiloop{
  N.~Berkovits,
  {\sl Multiloop amplitudes and vanishing theorems using the pure spinor formalism for the superstring,}
  JHEP {\bf 0409}, 047 (2004)
  [arXiv:hep-th/0406055].
}
\lref\BerkovitsTL{
 N.~Berkovits and C.~Mafra ,
  {\sl Equivalence of Two-Loop Superstring Amplitudes in the Pure Spinor and RNS Formalisms,}
  arXiv:hep-th/0509234.
}

\lref\BerkovitsTwoLoop{
  N.~Berkovits,
  {\sl Super-Poincare covariant two-loop superstring amplitudes,}
  arXiv:hep-th/0503197.
}
\lref\rfRusso{  J.~G.~Russo,
{\sl  Construction of SL(2,Z) invariant amplitudes in type IIB superstring theory,}
  Nucl.\ Phys.\ B {\bf 535}, 116 (1998)
  [arXiv:hep-th/9802090].
}
\lref\greengut{
M.B. Green and M. Gutperle, {\sl  Effects of D-instantons}, Nucl.\ Phys.\ B {\bf 498}, 195 (1997)
  [arXiv:hep-th/9701093].
}
\lref\gvtwo{M.B. Green and P. Vanhove, {\sl The low-energy
expansion of the one-loop type II superstring amplitude}, Phys.\ Rev.\ D {\bf 61}, 104011 (2000)
  [arXiv:hep-th/9910056].
}
\lref\rfWittenVarious{E. Witten, {\sl String Theory Dynamics in Various
    Dimensions}, Nucl.Phys. {\bf B443} (1995) 85, [arXiv:hep-th/9503124].
    }
\lref\rfSchwarzPower{J.H. Schwarz, {\sl The Power of M-theory}, Phys.Lett. {\bf
    367B} (1996) 97, [arXiv:hep-th/9510086].
}
\lref\rfAspinwall{P.S. Aspinwall, {\sl  Some Relationships Between Dualities in
    String Theory}, Nucl. Phys. Proc. Suppl. {\bf 46} (1996) 30,
    [arXiv:hep-th/9508154].
    }

\lref\rfBernDunbar{Z. Bern, L. Dixon, D.C. Dunbar, M. Perelstein and
    J.S. Rozowsky, {\sl On the Relationship between Yang-Mills Theory and
    Gravity and its Implication for Ultraviolet Divergences}, Nucl.Phys. {\bf
    B530} (1998) 401, [arXiv:hep-th/9802162];\hfill\break
 {\sl Perturbative Relationships Between QCD and Gravity and Some Implications},
  arXiv:hep-th/9809163.
  }
\lref\rfGreenGutperleKwon{M.B. Green, M. Gutperle and H. Kwon, {\sl
    Sixteen-Fermion and Related Terms in M-theory on $T^2$}, Phys.Lett. {\bf
    B421} (1998) 149, [arXiv:hep-th/9710151].
    }
\lref\rfGreenGutperleKwontwo{M.B. Green, M. Gutperle and H. Kwon,
{\sl Light-cone quantum mechanics of the eleven-dimensional
superparticle}, JHEP 9908:012,1999, [arXiv:hep-th/9907155].
}
\lref\rfGreenGutperleVanhove{M.B.Green, M. Gutperle and P. Vanhove, {\sl One-loop
    in Eleven Dimensions}, Phys.Lett. {\bf B409} (1997) 177, [arXiv:hep-th/9706175].
}
\lref\rfGreenVanhoveMtheory{M.B. Green and P. Vanhove, {\sl D-instantons,
    Strings and M-theory}, Phys.Lett. {\bf 408B} (1997) 122, [arXiv:hep-th/9704145].
 }
\lref\gvone{M.B. Green H. Kwon and P. Vanhove, {\sl Two Loops
    in Eleven Dimensions}, Phys. Rev. {\bf D61} 104010 (2000), [arXiv:hep-th/9910055].
    }

 \lref\rfRussoTseytlin{J.G. Russo and A.A. Tseytlin, {\sl One-loop four-graviton
    amplitude in eleven-dimensional supergravity}, Nucl.Phys. {\bf B508} (1997)
    245, [arXiv:hep-th/9707134].
    }
\lref\rfGreenSethi{M.B. Green and S. Sethi, {\sl Supersymmetry Constraint on
    Type IIB Supergravity}, Phys.Rev. {\bf D59} (1999) 046006,
    [arXiv:hep-th/9808061].
    }

\lref\rfGreenSchwarzWitten{M.B. Green, J.H. Schwarz and E. Witten, {\sl
    Superstring theory}, Cambridge University Press 1987.
    }
\lref\rfGreenSchwarz{M.B.~Green and J.H.~Schwarz, {\sl Supersymmetrical Dual
    String Theory. (II). Vertices and Trees}, Nucl.Phys. {\bf B198} (1982)
  252.
  }

\noblackbox
\baselineskip 14pt plus 2pt minus 2pt
\Title{\vbox{\baselineskip12pt
\hbox{DAMTP-2005-67}
\hbox{SPHT-T05/123}
}}
{\vbox{
\centerline{Duality and higher derivative terms in M theory
}
}}

\centerline{{\bf Michael B. Green}$^1$ and {\bf Pierre Vanhove}$^2$}
\medskip
\centerline{${}^1$ DAMTP,
Wilberforce Road, Cambridge CB3 0WA, UK
}
\centerline{\tt m.b.green@damtp.cam.ac.uk}
\medskip
\centerline{${}^2$Service de Physique Th{\'e}orique, CEA/DSM/PhT,
CEA/Saclay,}
\centerline{91191 Gif-sur-Yvette, France}
\centerline{\tt pierre.vanhove@cea.fr}
\bigskip

\medskip
\centerline{{\bf Abstract}}
Dualities of M-theory are used to determine the exact
dependence on the coupling constant
of the $D^6\, \calR^4$ interaction of the IIA and IIB
superstring effective action. Upon lifting to eleven dimensions
this determines the coefficient of the $D^6\, \calR^4$ interaction in
eleven-dimensional $M$-theory.   These results are obtained by considering
the four-graviton two-loop scattering amplitude
 in eleven-dimensional supergravity
compactified on a circle and on a two-torus -- extending
earlier results concerning lower-derivative interactions.
The torus compactification leads to an interesting
 $SL(2,\ZZ)$-invariant function of the complex structure of the torus
(the IIB string coupling) that satisfies a
Laplace equation with a source term on the fundamental domain of
moduli space.  The structure of this equation is in accord with general
supersymmetry considerations and
immediately determines  tree-level and one-loop
contributions to $D^6\calR^4$ in perturbative IIB string theory
that agree with explicit string calculations, and
two-loop and three-loop contributions that have yet to be obtained in string theory.
The complete solution of the Laplace equation contains
infinite series' of single $D$-instanton  and double
$D$-instanton contributions, in addition to the perturbative terms.
 General considerations
of the higher loop diagrams of eleven-dimensional supergravity suggest extensions of
these results to interactions of higher order in the low energy expansion.

\noblackbox
\baselineskip 14pt plus 2pt minus 2pt

\Date{PACS: 04.65.+e; 04.50.+h}

\listtoc
\writetoc

\newsec{Introduction}

The low-energy expansion of the effective action
of the type II superstring theories is an
infinite power series in $\alpha' = l_s^2$  (where $l_s$ is the string
distance scale)
consisting
of higher derivative interactions, which are strongly constrained by
maximal supersymmetry and $SL(2,\ZZ)$ invariance.   The leading term defines
the classical theory that contains the ten-dimensional Einstein--Hilbert action,
 $l_s^{-8} \int d^{10}x\, e^{-2\phi}\,\sqrt{-g} \, R$, together
with many other
interactions of the same dimension involving other fields.
These terms are uniquely specified
by imposing  IIB $N=2$  supersymmetry.

The absence of an off-shell superspace formalism for ten-dimensional extended supersymmetry indicates
that
the theory is very constrained, which makes it both difficult and interesting to determine the higher
derivative interactions.  Various duality and
supersymmetry arguments have been used to determine the form of some of the low order
terms \refs{\greengut,\rfGreenGutperleVanhove,\rfGreenGutperleKwon,\rfGreenGutperleKwontwo,\rfGreenSethi,
\SinhaZR}.
For example, the  first term in the derivative expansion  beyond the Einstein--Hilbert term
that contributes to four-graviton scattering  has the form
\eqn\rfour{l_s^{-2}\int d^{10}x\,
\sqrt{-g}\, e^{-\phi/2}\, Z_{\threeh}^{(0,0)}\, \calR^4\, ,}
in string frame.  The dilaton factor $e^{-\phi/2}$ is again  absent in Einstein frame. The symbol
$\calR^4$ denotes  a specific contraction of four Weyl tensors
 that arises from the leading behaviour in the low energy expansion
 of the four-graviton amplitude.  The function $Z_{3/2}^{(0,0)}(\Omega,\bar\Omega)$
 is a  modular form with
  holomorphic and anti-holomorphic weights $(0,0)$.  It is a function of the complex coupling
$\Omega = \Omega_1+ i \Omega_2$, where
 $\Omega_2 = e^{-\phi}$ and  $\Omega_1 = C^{(0)}$
(the Ramond--Ramond
zero-form).   There are very many other
interactions of the same dimension, that are related by supersymmetry to the $\calR^4$ interaction.
Many of these may be inferred by using a linearized on-shell superfield approximation in which the
interactions are given by integrals over sixteen Grassmann components, which is half the
dimension of the type II superspace.  However, it has not yet
 been possible to determine the full nonlinear
action at this order in the derivative expansion.

One of the methods for studying these higher order terms makes use of the
duality between eleven-dimensional supergravity
compactified on a two-torus in the limit of zero volume and ten-dimensional type II string theory,
which  was originally considered in the context of the classical theory
\refs{\rfAspinwall,\rfSchwarzPower},
based on properties of the supersymmetrised Einstein--Hilbert
action.  The IIB coupling constant is identified with the complex structure, $\Omega$, of the
two-torus, so the $SL(2,\ZZ)$ duality symmetry of the string theory originates from the geometric
invariance of supergravity under large diffeomorphisms of the torus.
Quantum corrections were considered in
\refs{\rfGreenGutperleVanhove,\rfGreenGutperleKwon,\rfGreenGutperleKwontwo},
where it was shown that the one-loop contributions to four-graviton scattering in
eleven-dimensional supergravity on $\calT^2$ determine the form of the
coefficient of the $\calR^4$ term,  $Z_{3/2}^{(0,0)}$, and its supersymmetric partners.
 This was generalized to the analysis of two-loop four-graviton scattering in
eleven-dimensional supergravity amplitudes compactified on $T^2$ in
\refs{\gvone}.  The leading term in the low energy limit
determined the dilaton dependent function, $Z_{5/2}^{(0,0)}(\Omega,\bar\Omega)$  of the
\eqn\twolop{l_s^2\int d^{10}x\, \sqrt{-g}\, e^{\phi/2}\,
Z_{\fiveh}^{(0,0)}\,  D^4 \calR^4\,}
 interaction, which is again expressed in string frame\foot{This symbolic notation
  indicates a term in which there are four (covariant) derivatives and four
  factors of the Riemann curvature.  The precise pattern of index contractions
  will be specified by the form of the amplitudes to be  calculated later.}.

The dilaton-dependent functions $Z_{3/2}^{(0,0)}$ and $Z_{5/2}^{(0,0)}$ in \rfour\ and \twolop\ are
non-holomorphic Eisenstein (or Epstein)
series that are special cases of the series
\eqn\eisendef{Z_{s}^{(w,w')}= \sum_{(m,n) \ne (0,0)}
{\Omega_2^{s}\over (m+ n \Omega)^{s+w} (m+ n \bar\Omega)^{s+w'}}\, .}
The modular weights $w$ and $w'$ are generally non-zero although they vanish in the case of interactions
$D^{4s-6}\calR^4$ terms (with $s=3/2, 5/2, \cdots$).  More generally, interactions have
$w = -w' = q/2$ where $q$ denotes
the $U(1)$ R-symmetry charge of the interaction under consideration.
For example, there is an
interaction of the form $\int d^{10}x\,\sqrt{-g}\, e^{-\phi/2}\, Z_{3/2}^{(12,-12)}
\, \lambda^{16}$ (where the dilatino $\lambda$ transforms with weights $(-3/4,3/4)$).
The series $Z_s^{(0,0)}$ is an eigenfunction of the Laplace operator on the fundamental
domain of $SL(2,\ZZ)$ with eigenvalue $s(s-1)$,
\eqn\lapold{\Delta_\Omega\, Z^{(0,0)}_s\equiv 4\Omega_2^2\, \partial_\Omega\partial_{\bar \Omega}\,
Z^{(0,0)}_s = s(s-1)\, Z^{(0,0)}_s\, .}
 It was shown in \refs{\rfGreenSethi} (in the case of the $\calR^4$
 interaction, which is the $s=3/2$ case) that this equation is a consequence of supersymmetry.
The series with non-zero $w$ and $w'$ is similarly an eigenfunction
of the Laplace operator with an $(w, w')$-dependent eigenvalue.

In the cases of relevance to this paper we always have $(w,w')=(0,0)$,
so we will drop the superscripts
from hereon.  For general values of $s$ $Z_s$
has the large-$\Omega_2$ (weak coupling)
expansion
\eqn\asymterm{\eqalign{ Z_s (\Omega,\bar \Omega) & = 2
\zeta(2s) \Omega_2^{s}  + 2\sqrt \pi  \Omega_2^{1 - s }
{\Gamma(s-\half)\zeta(2s -1)\over \Gamma(s)}\cr
&  + {2\pi^{s}\over \Gamma(s)}  \sum_{k\ne
0} \mu(k,s)
 e^{-2\pi(|k|\Omega_2 - i k \Omega_1)}
 |k|^{s-1}
 \left(1+{s(s-1) \over 4\pi |k| \Omega_2} +\dots\right)\, ,}}
where the last term comes from the asymptotic expansion of a modified
Bessel function and $\mu(k,s) = \sum_{d|k}1/d^{2s-1}$, as reviewed in appendix~A.
This expression
contains precisely two power behaved terms proportional to $\Omega_2^{s}$ and
$\Omega_2^{1 - s}$, which
should be identified with tree-level and $(s-1/2)$-loop term in the IIB string
perturbation expansion of the four graviton amplitude.
In addition, there is an infinite sequence of
$D$-instanton terms in $Z_{s}$, which have a characteristic
phase of the form $e^{2\pi i k\Omega}$,
where $k$ is the instanton number.

Thus, with $s=3/2$ (the $\calR^4$ term) there are tree-level and one-loop
terms, as well as the infinite series of $D$-instanton contributions.  The absence of
perturbative contributions to $\calR^4$ at two string loops has recently been confirmed by
 explicit evaluation of the two-loop string theory four-graviton amplitude
  \refs{\dhokerphong,\IengoPR} and to all orders in \refs{\BerkovitsMultiloop}.
  In the $s=5/2$ case (the $D^4R^4$ term determined in \refs{\gvone})
the non-holomorphic Eisenstein series $Z_{5/2}$ contains tree-level and two-loop perturbative
string theory contributions, as well as a sequence of $D$-instantons.  This agrees with the
absence of a one-loop contribution \refs{\gvtwo} and predicts that there
should be no perturbative terms beyond two loops, which has yet to be explicitly
verified by a string loop calculation.  The expression also predicts the value of
the two loop contribution to $D^4 \calR^4$ in type II string theory, which has recently been confirmed
by calculations of four-graviton scattering in \refs{\dgp,\dhokerphong,\dhokerphonglecture,\zhou} (see also
\refs{\BerkovitsTwoLoop,\BerkovitsTL}).

New features are expected to arise at the next  order in the low energy
expansion.   This is already clear from the form of the known tree-level contributions to higher derivative
interactions that come from the $\alpha'= l_s^2$
expansion of the tree-level four-graviton scattering amplitude summarized in
appendix A and  is proportional to
\eqn\treerough{\hat K^4 {1 \over l_s^8\, stu} \exp\left(\sum_{n=1}^\infty {2 \zeta(2n+1) \over
 2n+1}{l_s^{4n+2}\over 4^{2n+1}} (s^{2n+1} + t^{2n+1} + u^{2n+1})\right)\ ,
}
where $\hat K$ is the linearized Weyl curvature and  $s+t+u =0$.
At low orders the  coefficients are proportional to $\zeta$ functions that come from the
exponent in \treerough: $\zeta(3)$ in the case of
the $l_2^{-2}\, \calR^4$
interaction and $\zeta(5)$ in the case of $l_s^2 \, D^4 \calR^4$. The nonperturbative
extensions of these tree-level expressions  are given by $Z_{3/2}$ and $Z_{5/2}$, respectively.
However, the next term in the expansion
arises from the square of the exponent -- this is the term proportional to $\zeta(3)^2\, l_s^4 D^6 \calR^4$
with a coefficient that is the square of the $\calR^4$ coefficient.  The challenge is to determine
its nonperturbative extension.  One might simply guess  \refs{\rfRusso}
that it is proportional to $Z_{3/2}^2$, which contains the correct tree-level
term proportional to $\zeta(3)^2$, but this
is ruled out since it makes an
incorrect prediction for the one-loop contribution to $D^6 \calR^4$
\refs{\gvtwo}.  The correct
expression turns out to be much more subtle as we will see.

The objective of this paper is to extend the analysis of the dilaton dependence of higher
derivative interactions to the $D^6 \calR^4$ interaction.
This has the form (in string frame)
\eqn\dilrfour{l_s^4\int d^{10}x\,  \sqrt{-g}\, e^{\phi}\, \calE_{(\threeh,\threeh)} \, D^6\calR^4 \, ,}
where the function $\calE_{(3/2,3/2)}(\Omega,\bar\Omega)$
is a new $(0,0)$ modular form that depends on the complex coupling,
$\Omega$ (and the factor  $\Omega_2 = e^\phi$ disappears in Einstein frame).
  We will see that the function $\calE_{(3/2,3/2)}$ satisfies
a Laplace equation on moduli space with a source term,
\eqn\lapsource{\Delta_\Omega \calE_{(\threeh,\threeh)}
 =  12\calE_{(\threeh,\threeh)} - 6\,Z_{\threeh}Z_{\threeh}\,,}
and determine its solution.
These results will be obtained by expanding the two-loop supergravity amplitude
 compactified on $\calT^2$, which was considered in \refs{\gvone},  to first nontrivial order
  in the external momenta.

  The coupling constant dependence of the
  function $\calE_{(3/2,3/2)}$ encodes perturbative string tree-level, one-loop,
  two-loop effects  and three-loop effects (proportional to $\Omega_2^3$, $\Omega_2$,
$\Omega_2^{-1}$ and $\Omega_2^{-3}$, respectively),
together with an infinite number of $D$-instanton and double
$D$-instanton effects.  There are no other perturbative terms.
The $D$-instanton terms are absent in the IIA theory, which can be
obtained by compactification on a circle\foot{The four-graviton amplitudes in the
IIA and IIB theories are equal up to two loops -- they probably differ at higher orders due to the
contribution of odd-odd spin structures, which enter at three or more loops}.

The layout of the paper is as follows.
In section 2 we will review the two-loop calculation of \refs{\gvone},
where the $D^4\calR^4$
 term in the effective action were obtained by considering the low
energy expansion of the two-loop contribution to
eleven-dimensional supergravity compactified on $\calT^2$.  This  is greatly facilitated by the
observation in \refs{\rfBernDunbar} that the two-loop amplitude has a
  simple expression as a kinematic factor multiplying
a subset of the two-loop amplitudes of $\varphi^3$ {\it scalar} field
theory.   The kinematic factor is simply the linearized approximation to $D^4 \calR^4$
so that in \refs{\gvone} we simply set the external momenta in these scalar field theory
diagrams to zero in order to extract the effective $D^4 \calR^4$ interaction.
 The integral representations for  the loop diagrams compactified on a $n$-torus
were expressed
as  integrals over three Schwinger parameters.  These were
particularly easy to evaluate after redefining the  parameters so that the integrals were
expressed as integrals over the complex structure $\tau$ and volume $V$ of a two-torus.
Since the target space
of interest is also a torus (with complex structure $\Omega$ and volume $\calV$)
the value of the integral was obtained by considering mappings of
a torus into a torus.

Section 3  will be concerned with  the extension of the analysis to the
$D^6\calR^4$ interaction obtained by expanding
two-loop supergravity four-graviton amplitude to quadratic order in the external momentum.
Whereas in \gvone\ we had to carefully regulate a divergent integral, the term of relevance to
this paper is given by a finite integral.
In section 4 we will consider compactification  on a circle or radius $R_{11}$,
which is related to the IIA string theory.   In this case there is a single integer Kaluza-Klein
charge corresponding to the
discrete momentum in each loop.  After performing the continuous momentum
integrals the result is given as a sum over these integers.   The expression is converted
to a sum over windings ($\hm$ , $\hn$)
of the loops around the compact dimension, which isolates the divergence
in the zero winding sector ($\hm=\hn=0$).  The sum over non-zero windings
gives a finite expression that correctly reproduces the known string tree-level
coefficient proportional to $\zeta(3)^2 /R_{11}^6$.
As in the earlier cases \refs{\rfGreenGutperleVanhove,\gvone},
 the perturbative loop corrections of the IIA string theory are given by divergent expressions that are
 regulated by a cutoff in a manner that can be uniquely determined by relating them to the type IIB
 theory.  This is obtained by compactification on  $\calT^2$, which is considered in
 section 5.
After  performing the integration over the two nine-dimensional continuous loop momenta the
two-loop supergravity amplitude contribution to $D^6\calR^4$ will be
expressed as an  integral
over three Schwinger parameters and a sum over the four winding numbers ($(\hm_I, \hn_I)$ with
$I=1,2$) that correspond to windings of
either loop around either direction on the toroidal target space.  The terms we need
to keep in the limit that gives the
ten-dimensional type IIB theory  are those that survive when $\calV\to 0$ --
 this is the limit in which type IIB string theory
should be recovered.  These terms, which are proportional to $\calV^{-3}$,
are given by finite integrals and  unlike the terms discussed in \refs{\gvone} they
do not need to be regularized.

In section 5.1 we will  use an iterative procedure to evaluate these integrals, thereby leading
to an expression for  the dilaton-dependent coefficient, $\calE_{(3/2,3/2)}$, contained
in  \dilrfour.   This is given as a sum of two terms, $\calE_{(3/2,3/2)}= S+R$,
where $S$ is an infinite series and $R$ is an important remainder.

In section 5.2 we will show that $\calE_{(3/2,3/2)}$ has to
satisfy the inhomogeneous Laplace equation \lapsource\ on the
fundamental domain of on the
fundamental domain of $Sl(2,\ZZ)$ acting on $\Omega$.  We will argue that the general
structure of this equation
would be determined by a careful consideration of the conditions for  IIB
supersymmetry although we have not pursued this.
The series $S$ and the remainder $R$ do not separately  satisfy the Laplace
equation~\lapsource,
but the sum does.  In section 5.3 we will analyze properties of
 $\calE_{(3/2,3/2)}$ and calculate the coefficients of the perturbative terms.
 Extracting these  directly from the
solution is complicated but we can bypass this  by determining these coefficients
directly from properties of  the
Laplace equation.
 In particular, we will obtain the
values of the coefficients of the terms
proportional to $\Omega_2^3$, $\Omega_2$, $\Omega_{2}^{-1}$.
The  tree-level and one-loop
terms agree with those already known from string perturbation theory
and the value of the two-loop contribution is a new prediction since it has
not yet been extracted directly from string perturbation theory.
The evaluation of the three-loop contribution proportional to $\Omega_2^{-3}$ is more
subtle since it satisfies the homogeneous Laplace equation. In section 5.4 we will
determine the value of its coefficient using modular properties of the Laplace equation and
the fact that the $\calE_{(3/2,3/2)}$ is no more singular than $\Omega_2^3$.  Strikingly,
the value of the three-loop coefficient agrees with that of the three-loop contribution
to $D^6R^4$ in the type IIA theory that was contained in \refs{\gvone} (see also \refs{\rfRussoTseytlin}).
No other perturbative contributions arise beyond the three-loop term.

An important feature of
the two-loop and higher-loop terms in eleven-dimensional
supergravity  is that they  have overall
kinematic factors of the form $D^4\calR^4$, so that they do not give extra
contributions to  the one-loop $\calR^4$ term \refs{\rfBernDunbar}.  However,
the structure of supergravity Feynman diagrams is not sufficiently well understood
to know if diagrams with three or more loops will contribute to a renormalisation of
the $l_s^2\, D^4 \calR^4$ and $l_s^4\, D^6 \calR^4$ interactions.  The results of this paper indicate that
these interactions are completely accounted for by the two-loop  contributions
and should therefore not receive higher-order corrections.  This will be further discussed in
section 6.  We will give a dimensional argument that indicates that higher loop contributions
to eleven-dimensional supergravity cannot contribute to these  interactions.
Furthermore,  the general structure of the
three-loop diagrams will be used to constrain the form of the dilaton
dependence of the $l_s^6 \,D^8\calR^4$ and $l_s^8\, D^{10}\calR^4$ interactions.
Other comments concerning the systematics of higher order terms will also be made in section 6.

We end in section 7 with a summary that includes the evaluation of the
eleven-dimensional limit of the $l_s^4\, D^6\calR^4$ interaction.
This interaction, together with others of the same dimension,  are
the first nontrivial corrections to the eleven-dimensional
M-theory effective action after  $l_s^{-2}\, \calR^4$ (and other terms
of the same dimension) since the
$l_s^{2}\, D^4 \calR^4$ interaction vanishes in the eleven-dimensional
limit.



\newsec{Review of  two-loop supergravity and the $D^4\calR^4$ interaction}

Following \rfBernDunbar\ the two-loop four-graviton scattering amplitude
 in eleven-dimensional quantum supergravity has a very simple structure
 that can be expressed entirely in terms of a few scalar field theory diagrams.
\ifig\fthree{The `$S$-channel'
scalar field theory diagrams that contribute to the
two-loop four-graviton amplitude of eleven-dimensional supergravity.
(a)  The $(S,T)$ planar diagram, $I^P(S,T)$; (b) The $(S,T)$
non-planar diagram, $I^{NP}(S,T)$. }{\epsfbox{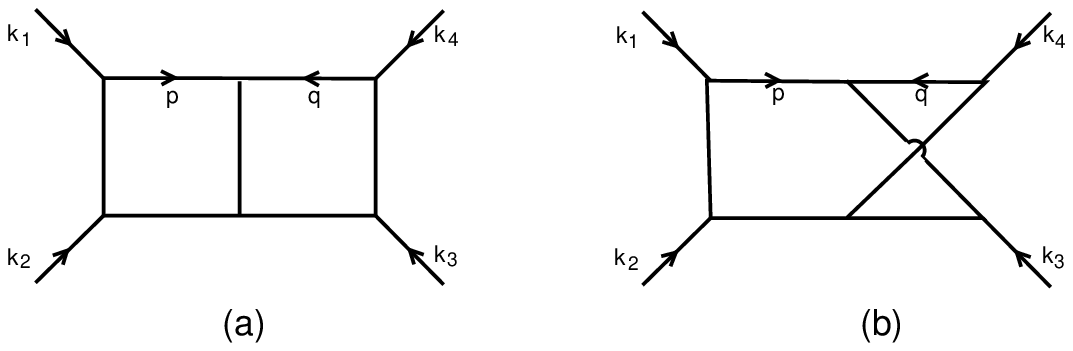}}

 The two-loop four-graviton amplitude\foot{Capital letters, $S$, $T$ and $U$ denote  Mandelstam
 invariants of the eleven-dimensional theory whereas lower case letters $s$, $t$ and $u$ denote
 Mandelstam invariants in the IIB string theory frame.},
$A^{(2)}_4(S,T,U)$,   is given in
terms of the sum of particular diagrams of $\varphi^3$
scalar field theory illustrated in \fthree.  These are the planar
diagram, $I^P(S,T)$, and the non-planar diagram, $I^{NP}(S,T)$, together
with the other diagrams obtained by permuting the external particles.
The complete expression for the amplitude is (with same conventions as in
\refs{\rfBernDunbar})
\eqn\eberndun{\eqalign{
A_4^{(2)} &=  \! i{\kappa_{11}^6\over (2\pi)^{22}}\; \hK\!
\left[S^2 I^{(S)} + T^2 I^{(T)} + U^2 I^{(U)}\right] \cr
& =\! i{\kappa_{11}^6\over (2\pi)^{22}}\; \hK\!
\left[S^2 \left(I^P(S,T)+\!  I^P(S,U)
+\! I^{NP}(S,T) +\! I^{NP}(S,U)\right)+ perms. \right],}}
where $\hat K$ is the kinematical factor given by the linearization of the $\calR^4$ term, and
 $perms$  signifies the sum of terms with permutations of $S,T$
and $U$ and
\eqn\sumall{ I^{(S)} (S,T,U)
={1\over2}\, \left(I^P(S,T) + I^P(S,U)+ I^{NP}(S,T) + I^{NP}(S,U)\right)\, , }
with analogous expressions for $I^{(T)}$ and $I^{(U)}$.
The  expression \eberndun\ has an overall factor of $\calR^4$ together with
four powers of the momentum multiplying the loop integrals which means
that these diagrams are much less divergent than they would
naively appear.  The loop integrals are given by
\eqn\eIP{
I^P(S,T)= \int d^{11}pd^{11}q {1\over p^2 (p-k_1)^2 (p-k_1-k_2)^2 (p+q)^2 q^2
  (q-k_3-k_4)^2 (q-k_4)^2}
}
and
\eqn\eINP{
I^{NP}(S,T)= \int d^{11}pd^{11}q {1\over p^2 (p-k_1)^2  (p+q)^2(p-k_1-k_2)^2 q^2
  (p+q+k_3)^2 (q-k_4)^2}\
}
which have ultraviolet divergences of order $(momentum)^8$ that will need to be
regularized.

In addition to these two-loop diagrams there is a contribution to the amplitude
from the one-loop triangle diagram
 in which there is one insertion of the  linearized one-loop
counterterm.  Together with the two-loop counterterm,  this
 gives  additional contributions  to the amplitude that are not relevant for the purposes
 of this paper.


\subsec{Evaluation of the two-loop amplitude on $\calT^n$}

Still following \gvone\
we shall now consider the leading contribution to the derivative
expansion arising from these two-loop diagrams when compactified
on $\calT^2$, which contributes
to the  $D^4 \calR^4$ interaction. For convenience
 our considerations will be restricted to situations in which the
polarization tensors and momenta of the gravitons are in
directions transverse to torus and covariantise the final result.
We will first be slightly more general and consider the
case of compactification on an
$n$-torus $\calT^n$ with metric $G_{IJ}$ and
volume $\calV_n$, in which case the  planar diagram with external momenta
$k_r$ $r=1,\dots,4$ is given by the expression,
\eqn\eintplan{\eqalign{
I^P(S,T) = &{1 \over l_{11}^{2n}\calV_n^2} \sum_{(m_I,n_I)} \int d^{11-n}p\;
d^{11-n}q\cr
& \int \prod_{r=1}^7 d\sigma_r \ e^{-\left[G^{IJ} \left(\sigma m_Im_J + \lambda
n_I n_J + \rho (m+n)_I (m+n)_J \right) + \sum_{r=1}^7 K_r \sigma_r \right]},
}}
where $I,J=1,2$ label the directions in $\calT^n$.  The vector $K_r$ is defined by
\eqn\kldef{K_r = (p,p-k_1,p-k_1-k_2,q,q-k_4,q-k_3-k_4,p+q),}
and
\eqn\schwindef{
\sigma=\sigma_1+\sigma_2+\sigma_3,\qquad \lambda=\sigma_4+\sigma_5+\sigma_6,
\qquad\rho=\sigma_7.}
The  non-planar diagram  is given by
\eqn\eIntegralNonPlanar{\eqalign{
I^{NP}(S,T) = & {1 \over l_{11}^{2n}\calV_n^2} \sum_{(m_I,n_I)}\int d^{11-n}p\;
d^{11-n}q\cr
&\quad  \int \prod_{r=1}^7 d\sigma_r \ e^{-\left[G^{IJ} \left(\sigma m_Im_J +
      \lambda n_I n_J + \rho (m+n)_I (m+n)_J \right) + \sum_{r=1}^7
  K^{\prime 2}_r \sigma_r \right]},\cr}
}
where
\eqn\nonplan{K'_r=(q, q-k_4, p, p-k_1, p-k_1-k_2, p+q, p+q+k_3),}
and
\eqn\schr{\sigma=\sigma_1+\sigma_2, \qquad\lambda=\sigma_3+\sigma_4+\sigma_5,
\qquad \rho=\sigma_6+\sigma_7.}

The loop momentum integrals are performed in the standard manner by
completing the squares in the exponent followed by gaussian
integration.  We are envisioning introducing some sort of  cutoff at
large momenta by imposing a lower limit to the  range of integration of
 the Schwinger parameters. The precise details will be clarified following
suitable changes of variables below. Ignoring these for now,
the resultant expressions for the planar
and non-planar loops are,
  \eqn\enonpll{\eqalign{
 &I^P(S,T) ={\pi^{11-n}\over l_{11}^{2n}\calV_n^2} \sum_{(m_I,n_I)}
 \int_0^\infty d\sigma d\lambda d\rho {\sigma^2\lambda^2\over
 \Delta^{11-n\over 2}} e^{-G^{IJ} \left(\sigma m_Im_J + \lambda n_I n_J + \rho
 (m+n)_I (m+n)_J \right)} \cr
& \int^1_0dv_2dw_2\int^{v_2}_0dv_1 \int^{w_2}_0dw_1
 e^{T{\sigma\lambda\rho\over
 \Delta}(v_2-v_1)(w_2-w_1)+S[{\sigma\lambda\rho\over \Delta}
 (v_1-w_1)(v_2-w_2)+\sigma v_1(1-v_2)+\lambda w_1(1-w_2)]},}}
and
\eqn\enonnon{\eqalign{
&I^{NP}(S,T) ={\pi^{11-n}\over l_{11}^{2n}\calV_n^2}
\sum_{(m_I,n_I)}\int_0^\infty d\sigma d\lambda d\rho {2\sigma\lambda^2\rho
\over \Delta^{11-n\over 2}} e^{-G^{IJ} \left(\sigma m_Im_J + \lambda n_I n_J +
\rho (m+n)_I (m+n)_J \right)}\cr
& \int^1_0du_1dv_1dw_2\int^{w_2}_0dw_1 e^{T{\sigma\lambda\rho\over
 \Delta}(w_2-w_1)(u_1-v_1) +S[{(\sigma+\rho)\lambda^2\over \Delta} w_1(1-w_2)
 +{\sigma\lambda\rho\over\Delta}( w_1(1-u_1) +v_1(u_1-w_2))]}}}
(where the variables $u_1$, $v_1$, $v_2$, $w_1$ and $w_2$ are rescalings of
$\sigma_i$).   These expressions can
be expanded in powers of $S,T$ and $U$ in order to determine their
 contributions to higher derivatives acting on $S^2\, \calR^4$.

The leading term in the low energy expansion (of order $S^2\, \calR^4$)
is obtained by
setting the external momenta to zero so
that $S$, $T$ and $U$ are set equal to zero in $I^P$ and
$I^{NP}$.
After summing these two zero-momentum contributions followed by a sum  over all the
diagrams required by Bose symmetrization the result is
\eqn\esumdi{
I^{P}(0)+I^{NP}(0)={\pi^{11-n} \over 3\;l_{11}^{2n}\calV_n^2}
\sum_{(m_I,n_I)}\int_0^\infty
d\sigma d\lambda d\rho\; {1\over \Delta^{7-n\over2}}
\; e^{- G^{IJ} \left(\sigma m_Im_J + \lambda n_I n_J + \rho (m+n)_I (m+n)_J
\right)},
}
which is symmetric in   the parameters
$\sigma,\lambda$ and $\rho$.  The integration  in \esumdi\
is  divergent for every value of $m^I,n^I$ when $\Delta \sim 0$, which requires
 at least  two of the parameters $\lambda,\rho,\sigma$ to approach
 zero simultaneously. The sums contribute additional  divergences, which makes this
 representation of the amplitude rather awkward to analyze.

As in the case of the one-loop amplitude \refs{\rfGreenGutperleVanhove} it is convenient to
analyze the divergences after
performing a Poisson resummation over the Kaluza--Klein modes, $m_I,n_I$,  which
transforms them into winding numbers, $\hm_I, \hn_I$, and also to
redefine the Schwinger parameters by,
\eqn\defsn{ \hsigma = {\sigma\over \Delta},\qquad \hlambda= {\lambda\over
\Delta}, \qquad  \hrho = {\rho\over \Delta},}
where
\eqn\deldef{\Delta = \sigma\lambda+\sigma\rho+\lambda\rho =
{\hdelta}^{-1}
=(\hsigma\hlambda+\hsigma\hrho+\hlambda\hrho)^{-1}.}
The amplitude \esumdi\ becomes (after a rescaling, $\hsigma\to \hsigma/\pi$)
\eqn\esum{
I^{P+NP}(0) ={\pi^7\over 3}
 \sum_{(m_I,n_I)} \int_0^\infty d\hsigma\, d\hlambda\, d\hrho\,
\hdelta^{1/2}\, e^{-\pi E_w} \ ,
}
where  the exponent is defined by
\eqn\eEw{
E_w(\hsigma,\hlambda,\hrho) = G_{IJ} \left(\hlambda \hm_I \hm_J +
\hsigma \hn_I \hn_J
+ \hrho (\hm+\hn)_I (\hm + \hn)_J
\right),
}
and is a function of the winding numbers.  The parameters $\hsigma$, $\hlambda$ and
$\hrho$ will be referred  to as `winding parameters'.
 The  classification of the
divergences is simplified in the winding number basis.
  For example, the sector in which all the
winding numbers vanish diverges at the end-point where
 all of the winding parameters reach their upper limits.
 This term  is independent of the metric $G_{IJ}$  and is the
primitive two-loop
 divergence.  There are many sectors that contribute to subleading
divergences. The simplest  examples are those  sectors
in which the winding numbers
conjugate to a particular winding  parameter  vanish.  In those cases
the integral  diverges
at the endpoint where that parameter reaches its upper limit,
 which gives a sub-leading divergence.  For example,
the $\hsigma$ integral diverges
 in the $\hat n_I=0$ sector and behaves as $\Lambda^3$  if
 $\hsigma$ is cut off at the value $\Lambda^2$ (that
 was introduced in order to cut off the one-loop winding parameter).
Sectors with less than $n$ vanishing winding numbers give
  non-divergent contributions which are independent of any cutoff.  This
  will be the situation for the interaction considered in the main part
  of this paper.

\ifig\ffour{ The domain of integration over the parameters $\tau_1$
and $\tau_2$,  bounded by the thick line, is the fundamental domain
of $\Gamma_0(2)$.}{\epsfbox{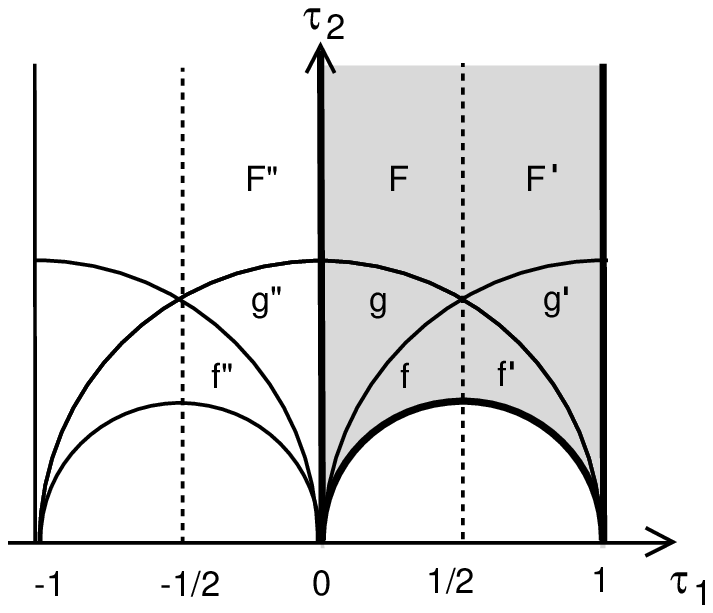}}
A more complete analysis of the integral
 is greatly  facilitated  by the observation that
the integrand possesses a secret $SL(2,\ZZ)$  symmetry that is not at
all apparent in the $\hlambda,\hrho,\hsigma$ parameterization.
This  symmetry is made  manifest by redefining  the integration variables in \esum\
 so that the parameters, $\hrho, \hlambda$ and $\hsigma$,
are replaced by the dimensionless volume, $V$, and complex
structure, $\tau = \tau_1 + i\tau_2$, of a two-torus, $\hat \calT^2$, defined
by
\eqn\modef{
\tau_1 = {\hrho\over \hrho+\hlambda}, \qquad \tau_2={\sqrt{\hat\Delta} \over
  \hrho+\hlambda},\qquad V=l_{11}^2\, \sqrt{\hDelta}\ .}
    The jacobian for the change of variables   from $(\hsigma,\hlambda,\hrho)$
    to  $(V,\tau)$ is
\eqn\jacob{d\hlambda d\hsigma d\hrho = 2 \,l_{11}^{-6} \,dV\, V^2\, {d^2\tau \over \tau_2^2},}
where $d^2\tau=d\tau_1 d\tau_2$. It is easy to see how the domain of
integration of the Schwinger variables translates into the integration domain
for $V$ and $\tau$.  The volume $V$ is integrated over $[0,\infty]$ and the
domain of integration of $\tau$ is the fundamental domain of the $\Gamma_0(2)$
sub-group of $SL(2,\ZZ)$ (the shaded region in~\ffour),
\eqn\domsdef{
\calF_{\Gamma_0(2)} = \left\{0\leq \tau_1\leq 1, \tau_2^2+\left(\tau_1 -
    {1\over2} \right)^2\geq {1\over 4} \right\},}
which consists of the sectors $F\oplus F'\oplus g\oplus g' \oplus f \oplus
f'$.  As is clear from the \ffour\ this domain covers precisely three copies
of  $\calF = F\oplus F^{\prime\prime}$, the fundamental domain of
$SL(2,\ZZ)$. More concretely, in terms of the conventional generators of
$SL(2,\ZZ)$:\foot{Which are the translation  $T$ :  $\tau \to \tau+1$ and  the
  inversion $S$:  $\tau \to -1/\tau$.}  region $g$ is mapped into
$F^{\prime\prime}$ by $S$;  region $g'$ is mapped into $F$ by $ST^{-1}$;
region $f$ is mapped into $F$ by $TS$; region $f'$ is mapped into
$F^{\prime\prime}$ by $T^{-1}ST^{-1}$; region $F'$ is mapped into
$F^{\prime\prime}$ by $T^{-1}$. Substituting the  change of variables \modef\ into  the integral \esum\
gives
\eqn\eintom{
I^{P+NP}(0) = {2 \pi^7\over l_{11}^8} \sum_{(m_I,n_I)} \int_0^\infty dV
V^3 \int_
{\calF_{\Gamma_0(2)}}  {d^2\tau\over \tau_2^2} e^{-\pi{V G_{IJ}\over
    l_{11}^2\tau_2} \left[(\hm + \tau \hn)^I(\hm + \bar \tau \hn)^J \right]} \
.
}

When the eleven-dimensional two-loop amplitude is compactified on a
  two-torus ($n=2$) of volume $\calV$ and complex structure $\Omega$
   the exponential factor
  \eEw\ can be written as
\eqn\eewt{
E = {\calV V\over \Omega_2 \tau_2} \left|(1\ \Omega)\, M \, (\tau\
  1)\right|^2 - 2\calV V \det M\ ,
}
where the metric on the two-torus is
\eqn\usug{
G_{IJ} \hm_I \hm_J =l_{11}^2\calV {|\hm_1+\hm_2\Omega|^2\over \Omega_2},}
and we have defined  a $2\times 2$ matrix $M$
\eqn\mdef{M = \pmatrix{\hm_1 & \hn_1\cr
                       \hm_2 & \hn_2 \cr}\, .}
In this case the expression \eintom\ becomes
\eqn\eun{
I_{D^4R^4} ={2\pi^7\over l_{11}^8}  \sum_{({\hm_I},{\hn_I})}
\int_0^\infty dV V^3 \int_{\calF}
{d^2\tau\over \tau_2^2}\,   e^{-\pi E}\, }

This integral, which resembles the integral that arises in evaluating a one-loop string amplitude in
a toroidally compactified space, was analyzed in detail in  \refs{\gvone}.
In particular, it was shown that
the coefficient of the $D^4 \calR^4$ interaction is determined by a one-loop sub-divergence,
which diverges like $\Lambda^3$.  This integral was regulated by introducing
the one-loop counterterm that had been determined previously from the analysis of the one-loop
$\calR^4$ term.  As a result the coefficient of the $D^4\calR^4$ term was found to be proportional to
$Z_{5/2}$, which is the $s=5/2$ case of  the non-holomorphic Eisenstein series \eisendef.

The tree-level part of $Z_{5/2}$ is proportional to $\zeta(5)\Omega_2^{5/2}$, and
the overall coefficient that emerged from the analysis in \refs{\gvone}
is precisely the one expected from the tree-level string four-graviton scattering amplitude.
Furthermore, the fact that $Z_{5/2}$ does not possess a $\Omega_2^{1/2}$ term implies that there
is no contribution to $D^4 \calR^4$ at one loop in string theory in ten dimensions\foot{
This is an example of how a contribution vanishes only after integration over the string moduli
(here the  one-loop modulus $\tau$).  This kind of phenomenon is expected to be the origin
of  non-renormalisation theorems for higher-derivative terms at higher-loop order.}
 as was verified in \refs{\gvtwo}.
The coefficient of the $\Omega_2^{-3/2}$ piece of $Z_{5/2}$ gave a prediction for the two-loop
contribution to $D^4\calR^6$ which has been verified by direct calculation in string
perturbation theory at two-loop order in \refs{\dgp}.

\newsec{$D^6\calR^4$ interaction from two-loop supergravity}

We now turn to consider the next term in the expansion of the two-loop
diagrams that contribute to $A_4^{(2)}$ in \eberndun.
This involves expanding  the integrals $I^P(S,T)$, $I^P(S,U)$, $I^{NP}(S,T)$
 and $I^{NP}(S,U)$ (\enonpll\
and \enonnon\ and the corresponding expressions with $T$ and $U$ interchanged)
to first order in the invariants $S$, $T$ and $U$.
This will give $(S^3+ T^3 + U^3) \calR^4$, which corresponds
to the terms of the form $D^6 \calR^4$ in the effective action,
\eqn\eVDsRf{
S_{D^6R^4}={l_{11}^5\over 96\cdot(4\pi)^{7}}\,\int d^{9}x \sqrt{-G^{(9)}} \,\calV\, h(\calV,\Omega,\bar\Omega)
\, D^6\calR^4\, .}
The function $h(\calV,\Omega)$ has an expansion in powers of $\calV$, which has the form
\eqn\hexps{h(\calV, \Omega,\bar\Omega) = {\pi^6\over4\calV^{3}}\,
\calE_{(\threeh,\threeh)}(\Omega,\bar\Omega) + \dots
\,,}
where $\dots$ indicates terms with higher powers of $\calV$, which are functions
of the cutoff but are negligible in the $\calV \to 0$ limit.
Using the dictionary in appendix B
 to those of type II string theory we can express the
 eleven-dimensional $D^6\calR^4$ action~\eVDsRf\ in terms of the
IIB string variables,
\eqn\eVDsRfII{
S^{(IIB)}_{D^6R^4}= l_{s}^5\,{\pi^6\over4\cdot 96\cdot(4\pi)^{7}}\, \int d^9x \sqrt{-g^{B}}\,
r_{B}\, e^{\phi^B}\, h(\calV,\Omega,\bar \Omega)\, \calV^{3}\, D^6R^4\,  ,
}
which has the finite $\calV \to 0$ limit given by \dilrfour,
\eqn\twobdr{
S^{(IIB)}_{D^6R^4}= l_{s}^4\, {\pi^6\over4\cdot 96\cdot(4\pi)^{7}}\,\int d^{10}x \sqrt{-g^{B}}\, e^{\phi^B}\,
\, \calE_{(\threeh,\threeh)}(\Omega,\bar\Omega)\,  D^6R^4\,  ,
}

\subsec{Expansion of two-loop integrals}

The result of expanding the sum of  diagrams contributing to $I^{(S)}$ in \sumall\
 to first order in the Mandelstam invariants
is (after integrating over $v_1,v_2,w_1,w_2$  and using $S+T+U=0$)
denote
\eqn\looppres{I^{(S)} =I^{P+NP}(0)+ { l_{11}^{2}\over 12} S \, I^{\prime}   + \cdots \, ,}
where
\eqn\exploop{
 I'  ={\pi^{11-n}\over 3\calV_n^2} \sum_{(m_I,n_I)}
 \int_0^\infty d\sigma d\lambda d\rho {(\lambda +\rho +\sigma)\Delta - 5\lambda\rho\sigma\over \Delta^{9-n\over 2}}
 e^{-G^{IJ} \left(\sigma m_Im_J + \lambda n_I n_J + \rho
 (m+n)_I (m+n)_J \right)}    .}

We now perform the Poisson resummations as before and
transform the integration variables from Kaluza--Klein charges to winding numbers
using \defsn.  This results in
\eqn\expeloop{
 I'  ={\pi^8\over3}\,  \sum_{(m_I,n_I)}
 \int_0^\infty d\hsigma d\hlambda d\hrho {\hlambda +\hrho +\hsigma- 5\hlambda \hrho \hsigma \hDelta^{-1}\over \hDelta^{1/2}}
 e^{-\pi G_{IJ} \left(\hsigma \hm_I \hm_J + \hlambda \hn_I \hn_J + \hrho
 (\hm+\hn)_I (\hm+\hn)_J \right)}    .}
Now we make the further transformations to the torus variables $\tau_1$, $\tau_2$ and $V$, defined
by \modef, which gives
\eqn\newexpo{
I' =   {2\pi^8}\,\sum_{(m_I,n_I)} \int_0^\infty dV
V^2 \int_{\calF_{\Gamma_0(2)}}   {d^2\tau\over \tau_2^2} \, \hat A\,e^{-\pi{V G_{IJ}\over
    l_{11}^2\tau_2} \left[(\hm + \tau \hn)^I(\hm + \bar \tau \hn)^J \right]} \
,
}
where
\eqn\adef{
\hat A =  {|\tau|^2 - \tau_1 + 1\over \tau_2}+5 {\tau_{1}(\tau_{1}-1)(|\tau|^2-\tau_{1})\over
\tau_{2}^3}\,.}

Although the function $\hat A$ is not invariant under $SL(2,\ZZ)$ it has simple transformation
properties.
Under both the inversion $S$
($\tau \to -1/\tau$) and under the translation $T$
($\tau \to \tau+1$), the function $\hat A(\tau_1,\tau_2)$ transforms into $\hat A(-\tau_1,\tau_2)$.
 Using this fact it
is easy to see how to map the regions $g$, $g'$, $f$, $f'$ and $F'$ into the fundamental domain
consisting of the regions $F''$ and $F$.   The result is that the integral \expeloop\ can be
replaced
by an integral over the fundamental domain of the form
\eqn\newexp{
I' = {2\pi^8}\,  \sum_{(m_I,n_I)} \int_0^\infty dV
V^2 \int_{\calF}   {d^2\tau\over \tau_2^2} \,  A\,e^{-\pi{V G_{IJ}\over
    l_{11}^2\tau_2} \left[(\hm + \tau \hn)^I(\hm + \bar \tau \hn)^J \right]} \
,
}
where
\eqn\adef{
A(\tau_1,\tau_2)  =   {|\tau|^2 - |\tau_1| + 1\over \tau_2}
+5 {(\tau_{1}^2-|\tau_1|)(|\tau|^2-|\tau_{1}|)\over
\tau_{2}^3}\,.}

It will prove important later that $A$ satisfies the Laplace equation\foot{As explained in
Appendix~C, this identity has to be understood as
a weak equality integrated over the fundamental domain $\calF$.}
\eqn\lapa{\Delta_{\tau} A=
\tau_2^2 (\partial^2_{\tau_1} + \partial^2_{\tau_2}) \, A
= 12\, A - 12\,\tau_2 \delta(\tau_1)}
%

\newsec{Compactification on a circle}

 A simple dimensional argument shows
that compactifying the four-graviton supergravity amplitude on a circle of radius $R_{11}$
(i.e., the case  $n=1$ in \exploop)  will give a finite term
of order $1/R_{11}^6$  which is to be identified with the IIA superstring
tree amplitude.   In this section we will find the coefficient of this term and see that it
is precisely the value expected from the direct calculation of the tree-level string amplitude.

In this case \newexp\ is given by
\eqn\newexpcirc{I' = {2\pi^8\over l_{11}^6}\,  \sum_{(\hm,\hn)'}
\int_0^\infty
dV V^2 \int_{\calF}   {d^2\tau\over \tau_2^2} \,  A\,e^{-{\pi E}} \
,}
where
\eqn\eedef{E = {{V R_{11}^2}\over
    l_{11}^2}\,{ |\hm + \tau \hn|^2\over\tau_{2}}\,,}
and $(\hm,\hn)'$ indicates that the value $(0,0)$ is omitted from the sum.

It is easy to see by changing the integration variable $V$ to   $V R_{11}^2$ that
$I' = a/R_{11}^6$, where $a$ is independent of $R_{11}$. From this it follows that
\eqn\derivione{\left(v^2 {\partial^2\over \partial v^2} + 2v {\partial \over \partial v }\right)
\, I' = 6I'\, ,}
where $v = R_{11}^2$.
 On the other hand, if we do not redefine the integration variable we have
\eqn\derivi{\eqalign{\left(v^2 {\partial^2\over \partial v^2}+ 2v {\partial \over \partial v }\right)
\, I' &= {2\pi^8\over l_{11}^6}\,  \sum_{(\hm,\hn)} \int_0^\infty dV
V^2 \int_{\calF}   {d^2\tau\over \tau_2^2} \,  A\,(\pi^2 E^2 - 2\pi E)\,  e^{-{\pi E}}\cr
& ={2\pi^8\over l_{11}^6}\,   \sum_{(\hm,\hn)}  \int_0^\infty dV
V^2 \int_{\calF}   {d^2\tau\over \tau_2^2} \,  A\,\Delta_\tau\,  e^{-{\pi E}}\, ,}}
where we have used the fact that
\eqn\exptr{\Delta_\tau \, e^{- {\pi E} }= (\pi^2 E^2 - 2\pi E)\, e^{-\pi E}\, ,}
as can be seen by simple manipulations.  Integrating \derivi\ by parts twice gives
\eqn\doubint{\left(v^2 {\partial^2\over \partial v^2}+ 2v {\partial \over \partial v }\right)I'
 = {2\pi^8\over l_{11}^6}\,  \sum_{(\hm,\hn)} \int_0^\infty dV
V^2 \int_{\calF}   {d^2\tau\over \tau_2^2} \, \Delta_\tau\, A\, e^{-\pi E}\, ,}
where the surface terms vanish so long as $\hm$ and $\hn$ are both non-zero.
Now we can use \lapa\ and \derivione\ to write this as
\eqn\final{6I' = 12I' - 12\, {2\pi^8\over l_{11}^6}\,    \sum_{(\hm,\hn)'} \int_0^\infty dV
V^2 \int_{1}^{\infty}   {d\tau_2\over \tau_2}\, \delta(\tau_{1}) \,
e^{-{\pi E}}\, .}
After making use of the symmetry of the integrand under $\tau_2\to 1/\tau_2$
$I'$ can be written as
\eqn\almfon{I' ={2\pi^8\over l_{11}^6}\,   \int_0^\infty V^2 dV \int_0^\infty {d\tau_2 \over \tau_2}\,
e^{-\pi {V R_{11}^2\over
    l_{11}^2\tau_2} (\hm^2 + \tau_2^2 \hn^2)}\, .}
We now change variables to $x$ and $y$, defined by
\eqn\xydefs{x = V\tau_2\, , \qquad y = {V\over \tau_2}\, ,}
so that $I'$ can be written as
\eqn\ress{\eqalign{I' &= {\pi^8\over  l_{11}^6}\,   \int_0^\infty dx\,  dy\, x^{1/2}y^{1/2} \, e^{-\pi {R_{11}^2\over
    l_{11}^2} (\hm^2 y +   \hn^2 x)}\cr
    & =  {\pi^6\over R_{11}^6}\, \zeta(3)^2 \, .}}
To this order in the momentum expansion the contribution of the
two-loop amplitude~\eberndun\ on a circle
is
\eqn\eAtwoExp{
A^{(2)}_{4} = {i\over2}\, {\kappa_{11}^6\over (2\pi)^22}\, {\pi^6\over l_{11}^8}\,
\hat K\,\left({\zeta(5)\over
R_{11}^5}(S^2+T^2+U^2)+{\zeta(3)^2\over6 R_{11}^6}l_{11}^2 (S^3+T^3+U^3)\right)\ ,
}
where we have included  the results of \refs{\gvone} for the $\zeta(5)\, D^4R^4$ term.
The relative normalization of $1/6$ between these terms agrees with the tree-level string
amplitude given in the appendix~A.  Since we showed in \gvone\ that the
normalization of the $\zeta(5)\, D^4R^4$ term also agrees with those of the lower order terms in
the series, we conclude that our two-loop calculation of the $D^6\, \calR^4$ term also agrees.

\newsec{Compactification on $\calT^2$}

As was seen in
the case of the $\calR^4$ term obtained from one loop in eleven dimensional supergravity in
\refs{\rfGreenGutperleVanhove}, compactification on a circle cannot determine the correct
regularization of divergent terms, which correspond to perturbative loop contributions to $\calR^4$
in  the IIA string action.  However,  one can determine these by compactifying on a two-torus
(the case $n=2$ in \exploop)
with volume $\calV$ and complex structure $\Omega$.  The limit
$\calV\to 0$ leads to a finite
term in the IIB action that contains the full non-perturbative
dependence on the complex
dilaton.  This contains specific tree-level and perturbative string
loop contributions as well as an infinite series
of $D$-instanton terms.   This determines the regulated IIA one-loop
contribution since
it is known that the four-graviton amplitude in the IIA theory only
receives perturbative contributions
that are equal to those
in the IIB theory, at least up to two loops.
In this section we will evaluate the leading term in the $\calV \to 0$
limit of the $\calT^2$
compactification of the $D^6 \calR^4$ interaction, which is the $n=2$
case of \exploop.

\subsec{Evaluation of integral}

We want to evaluate the integral \newexpo\ in the toroidal background
defined by the metric
\usug. It will also prove useful to define a generalization of $I'$,
labelled by an integer $p\ge 1$ (where $I_1 \equiv I'$ and the powers of
$l_{11}$ have been rescaled in $V$),
\eqn\eIp{
I_{p} = {2\pi^{8}}\,  \sum_{(\hm_{I},\hn_{I})'}\,
(\det M)^{2p-2}\,
\int_0^\infty dV V^{2p} \int_{\calF} {d^2\tau\over \tau_2^2}\, A\,
e^{-\pi E}\,
}
(where $(\hm_I,\hn_I)$  indicates that the values
$(\hm_1,\hm_2)=(0,0)$
and $(\hn_1,\hn_2)=(0,0)$ are omitted from the sum).
The fact that $E$ is proportional to $\calV V$ implies that $I_p$ has the form
\eqn\scaling{I_p = {2\pi^{8}}\, {{\cal I}_{p}\over
    \pi^{2p+1}\calV^{2p+1}}\, ,}
from which it follows, on the one hand, that
\eqn\resu{ \left(\calV^2 {\partial^2\over \partial \calV^2} + 2 \calV
  {\partial\over \partial \calV}
\right)\,{ I}_p = 2p(2p + 1)\, { I}_p}
and, on the other hand, the partial sums,
\eqn\ePartialSum{
I_{p}^{(\hm_{I},\hn_{I})}={{\cal I}^{(\hm_{I},\hn_{I})}_{p}\over \pi^{2p+1}\calV^{2p+1}}
 =\int_0^\infty dV V^{2p} \int_{\calF} {d^2\tau\over \tau_2^2}\, A\, e^{-\pi E}\,,
}
satisfy
\eqn\resut{\eqalign{ \left(\calV^2 {\partial^2\over \partial \calV^2} +
2 \calV {\partial\over \partial \calV}
\right)\,I_p^{({\hm_I},{\hn_I})}&=
\int_0^\infty dV V^{2p} \int_{\calF}
{d^2\tau\over \tau_2^2}\, A\, (\pi^2 E^2 - 2\pi E)\, e^{-\pi E}
\cr
&=
\int_0^\infty dV V^{2p} \int_{\calF}
{d^2\tau\over \tau_2^2}\, \left(A\, \Delta_\tau\,
 e^{-\pi E} + (2\pi \calV V\,
 \det M)^2\, e^{-\pi E}\right) \, .}}
Integrating the last expression by parts twice and combining it with \resu\ gives
\eqn\partss{\eqalign{2p(2p + 1)\, {\cal I}_p^{(\hm_{I},\hn_{I})} = &
 \pi^{2p+1}{\cal V}^{2p+1}\,\int_0^\infty dV V^{2p} \int_{\calF}
{d^2\tau\over \tau_2^2}\,  (\Delta_\tau\,A)\,
 e^{-\pi E} \cr
 & +   (2\det M)^2 \, {\cal I}_{p+1}^{(\hm_{I},\hn_{I})} + S_p^{\tau_2 =\infty}\, ,}}
where we have used
\eqn\laponexdp{\Delta_\tau \, e^{-\pi E}=
(\pi^2 E^2 - 2\pi E -( 2\pi {\calV} V \det M)^2)\, e^{-\pi E}\, ,}
and the fact that the surface terms vanish, apart from   contributions from the
boundary at $\tau_2 = \infty$, which are contained in the last term.
The vanishing of the surface terms follows from the fact that $A$ satisfies
the boundary conditions,
\eqn\bcs{\eqalign{&
A(\tau_1,\tau_2)|_{\tau_1=-\half} = A(\tau_1,\tau_2)|_{\tau_1=\half}\, ,\qquad
 A(\tau_1, \tau_2)|_{|\tau|=1} = A(-\tau_1, \tau_2)|_{|\tau|=1}\, , \cr
& \left.\partial_{\tau_1} A(\tau_1,\tau_2) \right|_{\tau_1 = \pm \half} = 0\,
,\qquad\qquad \qquad
\partial_{|\tau|} A(\tau_1,\tau_2)|_{|\tau|=1} = -\partial_{|\tau|}
A(-\tau_1,\tau_2)|_{|\tau|=1}\,.\,}}
The contribution from the
boundary at $\tau_2 = \infty$  is exponentially suppressed for terms with $\det M\ne 0$,
but does contribute  (and is divergent)  for terms in which $\det M =0$
(ie, for singular or degenerate orbits of $SL(2,\ZZ)$.)
Such terms need to be regulated as in \refs{\gvone}.
However, they have a higher power of the volume $\calV$, as follows from simple dimensional
analysis,
and are therefore suppressed in the IIB limit,
$\calV\to 0$ that we are considering, so $S_p^{\tau_2=\infty}$ will be ignored in the following.

 Substituting
$ \Delta_\tau\,A =12A - 12\tau_{2}\,\delta(\tau_1)$ in \partss\  gives
\eqn\partnew{
2p(2p +1)\,{\cal I}_p^{({\hm_I},{\hn_I})} = 12\, {\cal I}_{p}^{(\hm_{I},\hn_{I})}+ J_p^{({\hm_I},{\hn_I})}   + (2
 \det M)^2 \, {\cal I}_{p+1}^{({\hm_I},{\hn_I})} \, ,}
 where
 \eqn\jdef{
J_p^{({\hm_I},{\hn_I})} =  - 6 \, \pi^{2p+1}{\cal V}^{2p+1} \,
\int_0^\infty dV V^{2p}\left. \int_0^\infty {d\tau_2 \over \tau_2}
 e^{-\pi E}\right|_{\tau_1=0} \,.}
 The symmetry of the integrand under $\tau_2\to 1/\tau_2$
has been used to extend the integration range of the $\tau_2$ integral to the range $0\le \tau_2
\le \infty$.   Noting that
\eqn\taures{E|_{\tau_1=0} = {V\calV \over \Omega_2}\, \left[{|\hm_1+\hm_2\Omega|^2\over \tau_2}
+ \tau_2\, |\hn_1+\hn_2\Omega|^2\right]\, ,}
and setting $x = V\tau_2$ and $y=V/\tau_2$ gives
\eqn\finjr{\eqalign{J_p^{({\hm_I},{\hn_I})}  & = - 3 \, \pi^{2p+1} {\cal V}^{2p+1} \,
\int_0^\infty dx dy \, x^{p-\half}\, y^{p-\half}\, e^{-{\pi\calV \over \Omega_2}\left[
y |\hm_1+\hm_2\Omega|^2 + x |\hn_1+\hn_2\Omega|^2\right]}\cr
& = -3 \, \left(\Gamma\left(p+\half\right)\right)^2\,
{\Omega_{2}^{p+\half}\over |\hm_1+\hm_2\Omega|^{2p+1}}\,
{\Omega_{2}^{p+\half}\over |\hn_1+\hn_2\Omega|^{2p+1}}\,.}}
Summing~\partnew\ over $(\hm_{I},\hn_{I})$ we see that ${\cal I}_{p}$ satisfies the recursion relations
\eqn\eRecursion{\eqalign{
{\cal I}_{p}=  -3\, {(\Gamma(p+{1\over2}))^2\over (p-{3\over2})(p+2)}\,
\sum_{(\hm_{I},\hn_{I})'}\, (\det M)^{-3} \, Z_{p+\half}^{(\hm_{I})}Z_{p+\half}^{(\hn_{I})}
+{4\over (p-{3\over2})(p+2)}\,  {\cal I}_{p+1}
}}
where
\eqn\eZZ{\eqalign{
Z_{s}^{(\hm_{I})}&= {\Omega_{2}^{s}\over |\hm_{1}+\hm_{2}\Omega|^{2s}}\, ,\qquad
Z_{s}^{(\hn_{I})}= {\Omega_{2}^{s}\over |\hn_{1}+\hn_{2}\Omega|^{2s}}
\ }}
(so that $\sum_{(\hm_1,\hm_2)\ne (0,0)} Z_s^{(\hm_I)}=Z^{(0,0)}_s$).
The solution of this recursion relation gives
\eqn\eIone{
{\pi^6\over 4\calV^3}\, \calE_{(\threeh,\threeh)} =
I_{1}(\Omega,\bar\Omega)=  S(\Omega,\bar\Omega) + R(\Omega,\bar\Omega)\,,}
where $S$ is an infinite series,
\eqn\sdef{S(\Omega,\bar\Omega) =
{\pi^6\over 4\calV^3}\, \sum_{p=0}^{\infty}\,  c_{p} \,
\calZ_{(p+\threeh,p+\threeh)} \, ,}
with
 \eqn\cpdef{ c_{p}= {12\over\sqrt\pi} \, (1+2p)\,
   {\Gamma(p+{3\over2})\over \Gamma(p+4)}\, ,
}
and where we have introduced the generalized non-holomorphic Eisenstein series
\eqn\eCalZ{
\calZ_{(s,s')}= \sum_{ (\hm_{I},\hn_{I})'}
\, (\det M)^{s+s'-3} \, Z_{s}^{(\hm_{I})}Z_{s'}^{(\hn_{I})}
}
where  $\det M =\hm_{1}\hn_{2}-\hm_{2}\hn_{1}$ and the Kronecker delta restricts the sums
so that $\det M\neq 0$.   Notice that this sum vanishes when $s+s'$ is an even
number because of the cancellation between the $\det M>0$ and $\det M<0$
contributions.

The expression $R$ in \eIone\ is a remainder that is given by
\eqn\eRdef{\eqalign{
R(\Omega,\bar\Omega)& \equiv
- \lim_{p\to \infty}\, {2\pi^5\over \calV^3}\, {4\sqrt\pi\over
  \Gamma(p+{1\over2})\Gamma(p+4)}
 \, {\cal I}_{p+2}\cr
&= -\lim_{p\to\infty} \, {8\pi^{11\over2}\over
\calV^3}\,{\pi^{2p+5}\calV^{2p+5}\over
   \Gamma(p+{1\over2})\Gamma(p+4)}\,
\sum_{(\hm_{I},\hn_{I})'}\, (\det M)^{2p+2}\,  \int_{0}^\infty \, dV \,
V^{2p+4}\, \int_{\calF}\,
{d^2\tau\over\tau_{2}^2}\, A\, e^{-\pi E}\, .
}}
After integration over $V$ this becomes
\eqn\eRexpression{
R(\Omega,\bar\Omega)=  - \lim_{p\to\infty} \, {32\pi^5p\over \calV^3}\,
\sum_{(\hm_{I},\hn_{I})'}\,
\int_{\calF}\, {d^2\tau\over\tau_{2}^2}\, { (2\det M)^{2p+2}\,A(\tau) \over
\left({\left| (\hm_{1}+\hm_{2}\Omega)\tau + (\hn_{1}+\hn_{2}\Omega) \right|^2
\over \Omega_{2}\tau_{2}}-2\det M\right)^{2p+5}}\, .
}
Clearly if $\det M =0$ the series $S$ in \sdef\ truncates at the
first term and the remainder vanishes.  When $\det M\ne 0$, $R$ is
evaluated by taking the  $p\to \infty$ limit in  \eRexpression, in
which case   the integrand is dominated by the values of $\tau$ at
  which the magnitude
of the denominator,
\eqn\denomdef{\left|{\left| (\hn_{1}+\hn_{2}\Omega)\tau +
    (\hm_{1}+\hm_{2}\Omega)
 \right|^2\over \Omega_{2}\tau_{2}} -2\det M\right|\, ,}
is minimized.  It is easy to check that the  minima of this expression
arise when the
first term vanishes, i.e.,
at $\tau = \tau^0$, where
\eqn\tauzdef{\tau^0 = - {\hm_{1}+\hm_{2}\Omega\over
    \hn_{1}+\hn_{2}\Omega}\, ,}
so that
\eqn\tauztwo{\tau_2^0 = - {\det M\, \Omega_2\over |
    \hn_{1}+\hn_{2}\Omega|^2}\, .}
Setting $\tau = \tau^0+\epsilon$ and taking the limit $p\to \infty$
for values of $\hm^I$ and $\hn^I$ such that $\det M\ne 0$ gives
\eqn\rdeff{\eqalign{R(\Omega,\bar \Omega)& =- \lim_{p\to \infty}
{32\pi^5p\over\calV^3}\, \sum_{(\hm_{I},\hn_{I})'}\,{A(\tau^{0})\over
  |2\det M|^{3}(\tau_2^0)^2}\,
\int d^2\epsilon \, e^{-2p \log(1 - (\tau_2^0)^{-2} \,|\epsilon|^2 )}\cr
&=- {2\pi^6\over\calV^3}\,  \sum_{(\hm_{I},\hn_{I})'}\, (\det M)^{-3}\,
A\left(\hm_{1}+\hm_{2}\Omega\over \hn_{1}+\hn_{2}\Omega\right)\, .
}}

In summary, the complete solution naturally separates
 into contributions from the sector  with $\det M=0$
and the sector with $\det M\ne 0$,
\eqn\solfore{\eqalign{\calE_{(\threeh,\threeh)} = \calE^{\det M\ne 0}_{(\threeh,\threeh)}
+ \calE^{\det M=0}_{(\threeh,\threeh)} .}}
 In this expression we have
\eqn\eIfinala{\eqalign{
\calE^{\det M\ne 0}_{(\threeh,\threeh)}
= & \sum_{p=0}^{\infty}\,
     {12\over\sqrt\pi} \,
(1+2p)\, {\Gamma(p+\threeh)\over \Gamma(p+4)} \,
Z_{(p+\threeh,p+\threeh)}\cr
&
- 8
\,\sum_{(\hm_{I},\hn_{I})'} \, {\delta_{_{{\det M \neq0}}}\over  (\det M)^{3}}
A\left(\hm_{1}+\hm_{2}\Omega\over \hn_{1}+\hn_{2}\Omega\right)\, ,}
}
while
\eqn\detzero{\calE^{\det M=0}_{(\threeh,\threeh)} =
\sum_{(\hm_{I},\hn_{I})'}\, \delta_{_{\det M =0}}
Z^{(\hm_{I})}_{\threeh}Z^{(\hn_{I})}_{\threeh}\, . }

To conclude this subsection we note that
the sum of the remainder and the series can be reexpressed in a more compact
but rather formal manner by introducing the quantity
\eqn\eY{
Y_{(\hm,\hn)} = \det M^2\,
Z_1^{(\hm_I)}\, Z_1^{(\hn_I)}\equiv {\det M^2 \Omega_2^2\over |\hm_1 + \hm_2 \Omega|^2 \,
|\hn_1 + \hn_2 \Omega|^2}
= {\tilde\Omega^2_2\over \tilde\Omega^2_1 + \tilde\Omega^2_2}\, ,
}
where
\eqn\omegtransf{\tilde \Omega \equiv\tilde\Omega_{1}+i\tilde\Omega_{2}= M\cdot \Omega ={\hm_{1}+\hm_{2}
\Omega\over\hn_{1}+\hn_{2}\Omega}\, .}
Summing over $p$ in $S$ for $\det M\neq0$ gives
\eqn\eSrewrite{\eqalign{
S^{(\det M\neq0)}&= {\pi^6\over\calV^3}\, \sum_{(\hm,\hn)'}\,
(\det M)^{-3}\, \left(   20
Y^{-\threeh}_{(\hm,\hn)}- 18 Y^{-\half}_{(\hm,\hn)}
+{3\over 2} Y^{\half}_{(\hm,\hn)}\right)   \cr
&+{2\pi^6\over\calV^3}\, \sum_{(\hm,\hn)'}\,
{\sqrt{1-Y_{(\hm,\hn)}}\over(\det M)^3}\, \left(4 Y_{(\hm,\hn)}^{-\half}-
10Y_{(\hm,\hn)}^{-\threeh}\right)\, , }}
while expanding the
function $A$ in the remainder $R$ we find
\eqn\eRrewrite{\eqalign{R&={\pi^6\over\calV^{3}}\sum_{(\hm,\hn)'}\,
(\det M)^{-3}\, \left(8 \tilde\Omega_{2}^{-1} -
10 \tilde\Omega_{2}\, Y^{-2}_{(\hm,\hn)} +
8 \tilde\Omega_{2}\, Y_{(\hm,\hn)}^{-1}
- 10 \tilde\Omega_{2}^{-1} Y_{(\hm,\hn)}^{-1}\right)\cr
&-{2\pi^6\over\calV^3}\,\sum_{(\hm,\hn)'}\, {\sqrt{1-Y_{(\hm,\hn)}}\over(\det M)^3}\, \left(4 Y_{(\hm,\hn)}^{-\half}-
10 Y_{(\hm,\hn)}^{-\threeh}\right)\, .
}}
The $\hm^I$ and $\hn^I$ sums in the individual terms in
both \eSrewrite\ and \eRrewrite\ are divergent (even though the solution \eIfinala\ is not)
and must be defined by zeta function
regularization, so these expressions are rather formal.
Adding them together we see that the
 $\det M\ne 0$ sector simplifies  since  the square roots
 cancel with those in $R$.
As a result, the sum of \detzero\ and \eIfinala\ leads to the formal expression
\eqn\eIfinal{\eqalign{
\calE_{(\threeh,\threeh)}
&=6\calZ_{(\half,\half)}-72
\calZ_{(-\half,-\half)}+80\calZ_{(-\threeh,-\threeh)}
+64\calZ_{(0,-1)}-80\calZ_{(-2,-1)}\cr
& + \sum_{(\hm_{I},\hn_{I})'}\, \delta_{_{\det M= 0}} \,
Z_{\threeh}^{(\hm_{I})} \,Z_{\threeh}^{(\hn_{I})}\, .
}}

\subsec{Inhomogeneous Laplace Equation}\subseclab\secDif

In order to derive the Laplace equation satisfied by
$\calE_{(\threeh,\threeh)}$ we apply
 the Laplace operator $\Delta_{\Omega}=4\Omega_{2}^2 \,
\partial_{\Omega}\bar\partial_{\Omega}$
to  $I_1$, defined in \eIp.  We then use the fact that
\eqn\lapeq{\Delta_\Omega\,
e^{-\pi E} = \Delta_\tau \, e^{-\pi E}\,,}
followed by two integrations by parts (as in the step from \derivi\ to
\doubint) and the fact that
$\Delta_{\tau} A=  12 (A- \tau_{2}\delta(\tau_{1}))$.
The result is that  $\calE_{(\threeh,\threeh)}$ has to satisfy the
Laplace equation,
\eqn\eLaplaceEq{
\Delta_{\Omega}\calE_{(\threeh,\threeh)}= 12 \, \calE_{(\threeh,\threeh)} -
6\, \calZ_{(\threeh,\threeh)} \, ,
}
where  $\calZ_{(3/2,3/2)} \equiv Z_{3/2}\, Z_{3/2}$.

It is easy to check that the explicit expression \eIfinala,
which came from evaluating the two-loop
integral directly, indeed satisfies this equation.  For simplicity  we will here show
instead that the compact form \eIfinal\ satisfies the equation.
Firstly, note that the generalized series $\calZ_{(s,s)}$ in \eCalZ\
satisfies
\eqn\eZZDelta{
\Delta_{\Omega}\,\calZ_{(s,s')}=(s+s')(s+s'-1)\,\calZ_{(s,s')} - 4ss' \,
\calZ_{(s+1,s'+1)}
\, ,}
as shown in appendix~C.
Using this identity, the action of the $Sl(2,\ZZ)$ Laplacian on the solution~\eIfinal,
it is easily seen that
the first line of $\calE_{(3/2,3/2)}$ in~\eIfinal\ satisfies
\eqn\eLaplaceOne{\eqalign{
&[\Delta_{\Omega}-12] \left(6\calZ_{(\half,\half)}-72 \calZ_{(-\half,-\half)}
+80\calZ_{(-\threeh,-\threeh)}+64\calZ_{(0,-1)}-80\calZ_{(-2,-1)}\right)\cr
&= -6 \, \sum_{(\hm_{I},\hn_{I})'}\, (1-\delta_{_{\det M= 0}})\, Z_{\threeh}^{(\hm_{I})} Z^{(\hn_{I})}_{\threeh}\ ,
}}
and the second line of the $\calE_{(3/2,3/2)}$ in~\eIfinal\ satisfies
\eqn\eLaplaceTwo{\eqalign{
 \Delta_{\Omega}\!\!\!\!\!\! \sum_{(\hm_{I},\hn_{I})'}\!\!\!\!\!\!
\delta_{_{\det M=0}}\, Z^{(\hm_{I})}_{\threeh}Z^{(\hn_{I})}_{\threeh}
&=\sum_{(\hm_{I},\hn_{I})'}\!\!\!\!\!\!  \delta_{_{\det M= 0}}\, \left[6 Z^{(\hm_{I})}_{\threeh}Z^{(\hn_{I})}_{\threeh} - 9 Z^{(\hm_{I})}_{5\over2} Z^{(\hn_{I})}_{5\over2}\right]\cr
&= 6\, \sum_{(\hm_{I},\hn_{I})'}\!\!\!\!\!\!  \delta_{_{\det M= 0}}\, Z^{(\hm_{I})}_{\threeh}Z^{(\hn_{I})}_{\threeh}\ ,
}}
where we used in the last term that for $s+s'>3$ the generalized Eisenstein
series for  $\det M=0$ are vanishing.
By summing~\eLaplaceOne\ and~\eLaplaceTwo, one shows that the equation~\eLaplaceEq\ is satisfied.

It seems very likely that the Laplace equation \eLaplaceEq\ could also be derived
by supersymmetry considerations.
These would generalize the considerations of \refs{\rfGreenSethi}
where the Laplace eigenvalue equation
for the $l_s^{-2}\, \calR^4$ interaction
 \lapold\ was derived by considering the $O(l_s^6)$ modifications to
 the supersymmetry transformations that relate the
 $O(l_s^{-2})$
terms to the $O(l_s^{-8})$ classical action.  Supersymmetry similarly mixes
the $l_s^4\,
D^6R^4$ interaction with the classical action, which requires a new
$O(l_s^{12})$
modification to the
supersymmetry transformations.  However, a qualitative new feature
that first arises in this
 case is that the $O(l_s^6)$  supersymmetry transformations also mix   $l_s^{4}\, D^6R^4$
with   $l_s^{-2}\, \calR^4$ and other terms of the same order.    This explains
the generic origin of the inhomogeneous term, although have not studied this
 in any detail\foot{We are grateful to Savdeep Sethi for discussions on this issue}.

\subsec{Properties of $\calE_{(\threeh,\threeh)}(\Omega,\bar\Omega)$}

The solution \eIfinala\ is rather awkward to analyze directly so we
will here analyze those properties of $\calE_{(\threeh,\threeh)}$
 that can be seen directly from
the structure of the Laplace equation~\eLaplaceEq\ with rather little work.

To separate perturbative and non-perturbative contributions
we write $\calE_{(\threeh,\threeh)}(\Omega,\bar\Omega)$ in terms
of  a Fourier expansion of the form
\eqn\eIFourier{
\calE_{(\threeh,\threeh)}(\Omega,\bar\Omega)=
\tilde\calE_{(\threeh,\threeh)}^{(0)}(\Omega_{2}) +
\sum_{k\neq 0} \, \tilde\calE_{(\threeh,\threeh)}^{(k)}(\Omega_{2})\,
e^{2ik\pi \Omega_{1}}\, .
}
The dependence on $\Omega_1$ enters through the phase factor
$e^{2ik\pi \Omega_{1}}$,  that accompanies the non-zero mode.
This  is characteristic of a $D$-instanton contribution which
 comes from the double sum of $D$-instantons with charges $k_1$ and
 $k_2$, where $k_1+k_2 =k$.
There is a corresponding
exponentially decreasing coefficient, $\tilde\calE_{(3/2,3/2)}^{(k)}$,
that should behave as
$e^{-2\pi  (|k_1|+ |k_2|)\Omega_2}$ at weak coupling ($\Omega_2\to \infty$).
The zero mode, $\tilde\calE_{(3/2,3/2)}^{(0)}$, contains the piece that is a power-behaved
function  of the inverse string coupling constant,
$\Omega_2$ which is interpreted as a perturbative string contribution.
There will also be an exponentially decreasing contribution to the zero mode piece, which
is interpreted as a double $D$-instanton contribution in which the instanton charges are
equal and opposite in sign ($k_1=-k_2$).

\medskip
{\it (a) The zero mode contribution $ \tilde\calE_{(\threeh,\threeh)}^{(0)}(\Omega_{2})$}
\hfill\break\noindent
The zero mode  in \eIFourier\ satisfies the equation
\eqn\zeromode{\eqalign{
(\Omega_{2}^2\, & \partial^2_{\Omega_{2}} - 12)\,
 \tilde\calE_{(\threeh,\threeh)}^{(0)}(\Omega_{2})=\cr
&- 6\,
    \left((2\zeta(3)\Omega_{2}^{\threeh}+4\zeta(2)\Omega_{2}^{-\half})^2
+ (8\pi)^2\Omega_{2}\sum_{k\neq 0} k^2\mu^2(k,\threeh)\, \calK_{1}^2(2\pi|k|\Omega_{2})\right)
\, ,}}
where the right-hand side cones from a Fourier expansion of $Z_{3/2}^2$.
The factor $(2\zeta(3)\Omega_{2}^{3/2}+4\zeta(2)\Omega_{2}^{-1/2})^2 $
comes from the square of the zero mode  of $Z_{3/2}$ (defined by the first line in  \asymterm\
with $s=3/2$)
whereas the term involving the square of Bessel functions $\calK_1^2$ comes from the modes with non-zero $k$, which arise as a
 sum over $D$-instanton anti $D$-instanton pairs with $k_1=-k_2$ and $|k_1|+ |k_2| = k$.  The quantity
 $\mu(k,3/2)=\sum_{d|k}\, d^{-2}$  is the D-instanton measure factor \refs{\greengut}.

Consider first  the solution for the perturbative part of
$\tilde\calE_{(3/2,3/2)}^{(0)}$, which is a sequence of power-behaved terms.
The general solution for the power behaved terms that satisfy   \zeromode\
is
\eqn\eIonepert{
\tilde \calE_{(\threeh,\threeh)}^{(0)\, pert}=
4 \zeta(3)^2\, \Omega_{2}^{3}+8\zeta(3)\zeta(2)\Omega_{2}+
{48\over 5}\zeta(2)^2 \Omega_{2}^{-1} + \alpha \,
\Omega_{2}^{4}+\beta\, \Omega_{2}^{-3} \, ,
}
where the coefficients $\alpha$ and $\beta$ are not determined directly
by \zeromode\ because the terms $\Omega_2^4$ and
$\Omega_2^{-3}$  individually
satisfy the homogeneous equation.
It is easy to see from the solution that the $\Omega_2^4$
term is absent, so $\alpha=0$
(it would obviously not have made sense for $\alpha$ to be non-zero
since a $\Omega_2^4$ term would be more singular than the
tree-level $\Omega_2^3$ term).
The coefficient $\beta$
represents a three-loop contribution in string perturbation
theory.  The value of  $\beta$ requires separate
discussion and is the subject of the next subsection.

The remaining coefficients in \eIonepert\ are  determined
directly by \zeromode.
These correspond to the tree-level, one-loop and two-loop
contributions to the $D^6R^4$ interaction.
The leading $\Omega_2^3$ term in \eIonepert\
represents the tree-level contribution and has precisely the expected
coefficient that matches the  string tree-level calculation that is
reviewed in appendix~A.
The coefficient of the one-loop term proportional to $\Omega_2$ is exactly
one half of the value that
would have arisen if we had assumed that $\calE_{(3/2,3/2)}$ were given by $Z_{3/2}^2$,
as in \refs{\rfRusso}.  But
the   analysis of the one-loop four-graviton scattering
amplitude in type II string theory in \gvtwo\ determined
that the correct value for this coefficient
is one half of the value
contained in $Z_{3/2}^2$, so our value agrees with the correct value.
In addition, the expression \eIonepert\ includes the two-loop term
$48\zeta(2)^2\, \Omega_2^{-1}/5$.
Since this has not yet been calculated in string perturbation theory
this gives a prediction that should be calculable by an extension of
\refs{\dgp,\BerkovitsTwoLoop,\dhokerphong,\dhokerphonglecture,\zhou} to include the first non-leading term in the expansion of the
string-theory two-loop term in powers of the external momenta.

In addition to the power behaved piece, $\tilde \calE_{(3/2,3/2)}^{(0)}$
contains an exponentially
decreasing piece that comes from charge-$\hat k$ $D$-instanton and
charge-$(-\hat k)$ anti $D$-instanton pairs that contribute to the sector with
$k=k_1+k_2=0$.  This can again be discovered
directly from the solution  or else by setting $\hat k=k_1=-k_2$ in the
following analysis of the more general charge-$k$ sectors, which contain only
non-perturbative contributions.

\medskip

{\it (b) Non-perturbative terms,
$\tilde\calE_{(\threeh,\threeh)}^{nonpert\, (k)}(\Omega_{2})$}
\hfill\break\noindent
Having determined the perturbative contributions, the remaining contributions involve single
charge-$k$ $D$-instantons or pairs of $D$-instantons with net charge $k$.

Expanding \eLaplaceEq\ in Fourier modes gives an equation for each mode of the form
\eqn\eallmode{\eqalign{
[\Omega_{2}^2\, & (\partial^2_{\Omega_{2}}- 4\pi^2 k^2 )- 12]
 \tilde\calE_{(\threeh,\threeh)}^{nonpert\, (k)}(\Omega_{2})\cr
&=- 384\pi^2\,
 \Omega_{2}\sum_{k_1\ne 0,k_2\ne0\atop k_1+k_2=k} |k_1\, k_2|\, \mu(k_1,\threeh)\,\mu(k_2,\threeh)\, \calK_{1}(2\pi|k_1|\Omega_{2})\,
 \calK_{1}(2\pi|k_2|\Omega_{2})\cr
 & -96\pi\, \left(2\zeta(3)\Omega_{2}^{\threeh}+ 4\zeta(2)\Omega_{2}^{-\half}\right)
 \sum_{k_{1}\neq0}\, |k_{1}|\,
 \mu(k_{1},\threeh)\, \calK_{1}(2\pi |k_{1}|\Omega_{2})
\, ,
 }}
 Using the asymptotic form for the modified Bessel function $\calK_1(z) \sim \sqrt{\pi/2z} \, e^{-z}$, the large-$\Omega_2$
limit of the solution is easy to determine.  For a general value of $k=k_1+k_2$ it has the form
\eqn\eGeneralWeight{
\sum_{k_1} P_{k_1}(\Omega_2)\, e^{-2\pi (|k_1|+|k- k_1|)\Omega_{2}}\, e^{2\pi i k \,\Omega_{1}}\, ,}
where the functions $P_k\sim \Omega_2^{-p_k}$ with positive $p_k$.   When $k_1$
and $k_2$ $(= k-k_1)$ both have the same sign the action is equal to the charge ($|k_1|+|k- k_1| = k$).
But otherwise the action is less than the charge.  In particular, there is a $k=0$ contribution to
$\calE_{(3/2,3/2)}^{(0)}$ due to  $D$-instanton---anti $D$-instanton pairs, mentioned
above, that has the form
\eqn\paircont{- 64\pi^2\,\sum_{k} \, |\hat k|\mu(\hat k,\threeh)^2\,
\left({1\over 4\pi |\hat k|\Omega_2}
+\cdots\right)\, e^{-4\pi |\hat k|\Omega_{2}} \, .}

\subsec{The three-loop term}
We will now determine the three-loop coefficient,  $\beta$, of the $\Omega_2^{-3}$ term  
 in~\eIonepert.  First  we should note that a general solution of the Laplace
equation \lapsource\ can be written as the
sum of a particular solution
 and  a multiple of $Z_4$, which is  the solution
of the homogeneous Laplace equation, $\Delta\, Z_4 = 12\, Z_4$.
Recall also that  $Z_{4}= \sum_{(m,n)\neq (0,0)} \Omega_{2}^4/|m+n\Omega|^{8}$
has the large-$\Omega_2$ expansion
\eqn\zfourexp{Z_{4}=2\zeta(8)\Omega_{2}^{4}+{5\pi\over 8}\, \zeta(7)\,
 \Omega_{2}^{-3}  + \dots\, }
 where $\dots$ denotes exponentially suppressed terms.  However, the
 special solution $\calE_{(3/2,3/2)}$ that we  obtained from the two-loop
supergravity expression is known not to have a
 $\Omega_2^4$ piece  ($\alpha=0$ in \eIonepert), so that
 the coefficient of $Z_4$ in the general solution must be zero.
The question remains as to whether  $\calE_{(3/2,3/2)}$
contains a $\beta \,\Omega_2^{-3}$ term.

To study this we multiply the left-hand
and right-hand sides of the inhomogeneous
Laplace equation \lapsource\ by the Eisenstein series
$Z_4$
 and integrate over
a fundamental domain of $\Omega$. Since the relevant integrals diverge at
the boundary $\Omega_2\to \infty$, we will introduce a cut-off at
 $\Omega_2 = L$ and consider the $L\to \infty$ limit.
 Denoting the cut-off fundamental domain by
 $\calF_L$, the resulting equation is
\eqn\lapzfour{\int_{\calF_L} {d^2\Omega\over\Omega_{2}^2} \,
Z_{4}\,\Delta\calE_{(\threeh,\threeh)}
 =  12\int_{\calF_L} {d^2\Omega\over\Omega_{2}^2} \,
 Z_{4}\,\calE_{(\threeh,\threeh)} -
 6\int_{\calF_L} {d^2\Omega\over\Omega_{2}^2} \, Z_{4}\,Z_{\threeh}^2\,.}
Integrating the left-hand side by parts and using the fact that
 $\Delta\, Z_4 = 12\, Z_4$, gives
\eqn\eZcE{\eqalign{
\int_{\calF_L} {d^2\Omega\over\Omega_{2}^2} \,
 Z_{4}\Delta\calE_{(\threeh,\threeh)}
&=  \int_{\calF_L}
 {d^2\Omega\over\Omega_{2}^2} \Delta Z_{4} \, \calE_{(\threeh,\threeh)}+
 \int_{-\half}^{\half}\!\!\! d\Omega_{1}\, \left.\left(Z_{4}
 \partial_{\Omega_{2}}\calE_{(\threeh,\threeh)}-
 \partial_{\Omega_{2}}
 Z_{4}\,\calE_{(\threeh,\threeh)}\right)\right|_{\Omega_{2}=L}\cr
 &= 12\int_{\calF_L} {d^2\Omega\over\Omega_{2}^2}  \,
 Z_{4}\,\calE_{(\threeh,\threeh)} +
 \int_{-\half}^{\half}\!\!\! d\Omega_{1}\, \left.\left(Z_{4}
 \partial_{\Omega_{2}}\calE_{(\threeh,\threeh)}-
 (\partial_{\Omega_{2}}
 Z_{4})\,\calE_{(\threeh,\threeh)}\right)\right|_{\Omega_{2}=L} \, .
}}
Comparing \eZcE\ with \lapzfour\ we see that
\eqn\ebb{
\int_{-\half}^{\half}d\Omega_{1}\, \left.\left(Z_{4}
\partial_{\Omega_{2}}\calE_{(\threeh,\threeh)}- (\partial_{\Omega_{2}} Z_{4})\,
\calE_{(\threeh,\threeh)}\right)\right|_{\Omega_{2}=L\to\infty}
= -6 \int_{\calF_{L}} {d^2\Omega\over\Omega_{2}^2} Z_{4} \, Z_{\threeh}^2 \, .}

The left-hand side of this equation is simply a surface time that is easy to evaluate
\eqn\surfter{\eqalign{\int_{-\half}^{\half}d\Omega_{1}\,& \left.\left(Z_{4}
\partial_{\Omega_{2}}\calE_{(\threeh,\threeh)}- (\partial_{\Omega_{2}} Z_{4})\,
\calE_{(\threeh,\threeh)}\right)\right|_{\Omega_{2}=L\to\infty}
=  \cr
& -\zeta(8)\left(8\zeta(3)^2 \, L^6+48 \zeta(3)\zeta(2)
  L^4+ 96\zeta(2)^2\, L^2+14 \, \beta\right) .}}
The right hand-side of~\ebb\ may be evaluated by unfolding the integral\foot{This
 is the standard Rankin-Selberg trick which states that one can unfold
integrals of Poincar\'e series onto the strip  \refs{\Terras}
$$
\int_{\calF} {d^2\tau\over\tau_{2}^2} \sum_{\gamma\in\Gamma_{\infty}\backslash PSl(2,\ZZ)} \psi(\gamma\cdot\tau)\, f(\tau)
= \int_{0}^{\infty} {d\tau_{2}\over \tau_{2}^2}
 \psi(\tau_{2})\, \int_{-\half}^{\half} f(\tau) d\tau_{1}\ ,
$$
where $\Gamma_{\infty}=\pmatrix{1&n\cr 0&1}$ and
$\calF= PSl(2,\ZZ)\backslash{\cal H}$ is the fundamental domain for
 $Sl(2,\ZZ)$ and ${\cal H}=\{\tau =\tau_{1}+i\tau_{2}| \tau_{2}>0\}$ is the upper half-plane.
}
 onto the strip using  $Z_{4}=2\zeta(8)\sum_{\gamma\in Sl(2,\ZZ)}\,
 \Im {\rm m}(\gamma\cdot\Omega)^4$ and the fact that $Z_{3/2}^2$ is modular invariant,
 which gives
 \eqn\eunfol{\eqalign{
{1\over2\zeta(8)} \int_{\calF_{L}} {d^2\Omega\over\Omega_{2}^4} Z_{4} Z_{\threeh}^2& =\int_{0}^{L} {d\Omega_{2}\over\Omega_{2}^2} \, \Omega_{2}^4 \, \int_{-\half}^\half d\Omega_{1}\, Z_{\threeh}^2\cr
 &= {2\over3}\zeta(3)^2 \, L^6+ \zeta(3)\zeta(2) \, L^4+ 2\zeta(2)^2\, L^2 \cr
 &+(8\pi)^2\,\int_{0}^{L} d\Omega_{2}\,\Omega_{2}^3\, \sum_{k\neq0} k^2 \mu(|k|,\threeh)^2 \,
 \calK_{1}^{2}(2\pi |k| \Omega_{2})\, .
  }}
Using the integral representation for the Bessel function given in appendix~A we find  that
\eqn\eThreeLoop{\eqalign{
\beta&={384\pi^2\over 7} \int_{0}^{\infty} d\Omega_{2}\,\Omega_{2}^3\,
\sum_{k\neq0} k^2 \mu(|k|,\threeh)^2 \, \calK_{1}^{2}(2\pi |k| \Omega_{2})
= {32\over 7 \pi^2} \sum_{k\geq 1} {\mu(k,\threeh)^2\over  k^2}\, ,
}}
which gives a non-zero value for the three-loop term. Recalling that $\mu(n,s)= \sum_{m|n} \, n^{1-2s}$ and using an  identity by Ramanujan  quoted in
 \refs{\Apostol}
\eqn\eDivisor{
\sum_{n=1}^{\infty} {\mu(n,s)\mu(n,s')\over n^{r}} = {\zeta(r)\zeta(r+2s-1)\zeta(r+2s'-1)\zeta(r+2s+2s'-2)
\over \zeta(2r+2s+2s'-2)}
}
we find that the three-loop coefficient has the value
\eqn\eBeta{
\beta = {16\over 189}\pi^2\, \zeta(4)\ .
}

We cannot compare this predicted coefficient with  perturbative string theory since
there are no explicit  three-loop results.  However, this number is in complete agreement
with our earlier calculation of the three-loop coefficient in type IIA string theory that
is contained in \refs{\gvone} (see also \refs{\rfRussoTseytlin}). There it was shown
that the one-loop
four-graviton amplitude amplitude of eleven-dimensional supergravity
compactified on a two-torus  gives rise to a series of
higher-derivative terms  in the nine-dimensional type IIA effective action of the form
\eqn\estringa{
\eqalign{   A_4^{(1)}=& (4 \pi^8 l_{11}^{15}r_A^{-1}) \;\hK\,
r_A\left[
 2\zeta(3)e^{-2\phi^A}+ {2\pi^2\over 3 r_A^2} + {2\pi^2\over 3}
 -8\pi^2 r_A\; l_s (-\calw)^{\half}\right.\cr
 &\left. +8\pi^{3\over 2}\sum_{n=2}^\infty \left(\Gamma(n-\half)\zeta(2n-1)
  {r_A^{2(n-1)}\over n!}(l_s^2 {\calw})^n
\right.\right. \cr & \left.\left. + \sqrt \pi \Gamma(n-1)
 \zeta(2n-2) {e^{2(n-1)\phi^A}\over n!}(l_s^2{\calw})^n \right)
 \right]   + {\rm non-perturbative}, \cr}}
 where
 \eqn\eWn{
 (\calw)^n  = (\calgst)^n+(\calgtu)^n+ (\calgus)^n\, ,}
 and
 \eqn\gdefs{
( \calgst)^n  = \int_0^1 d\omega_{3} \int_0^{\omega_{3}} d\omega_{2} \int_0^{\omega_{2}} d\omega_{1}\,
  \left(s\omega_{1}(\omega_{3}-\omega_{2})+t (\omega_{2}-\omega_{1})(1-\omega_{3})\right)^n\, .
 }
The terms in the third line of \estringa\
give higher-loop contributions to the ten-dimensional effective action
of the type IIA theory.
 The term with $n=2$ gives the  two-loop $D^4R^4$ term  in the IIA theory
 that matches the  same term in the type IIB theory that was the subject of \refs{\gvone}.
 The term with $n=3$ in the third line of \estringa\
contributes to the three-loop  $D^6R^4$ term in the IIA theory and has the value
\eqn\eIIA{
S_{D^6R^4}^{(IIA)}=l_{s}^4\, {1\over  4\cdot 96\cdot(4\pi)^{7}}{16\over 189}\,\pi^2\,\zeta(4) \, \int d^{10}x \sqrt{-g^{A}}\, e^{4\phi^{A}}\, D^6 R^4\ .
}
Including the absolute normalisation this type IIA expression and
the type IIb expression~\twobdr,
we find a perfect match between  the two values for the three-loop
coefficient for the $D^{6}R^{4}$ in superstring theory.
 This may be of interest since there seems to be no reason,
a priori, for the three-loop four-graviton amplitudes to be equal in the two theories
as pointed out in\foot{The amplitudes
 only differ in the sign of the odd-odd spin structures. At one and two loops
the odd-odd spin structures vanish, so the amplitudes must be equal, but they
 need not vanish at
three or more loops.} \refs{\gvone}.

\newsec{Higher Loops and Higher Order Interactions}

The results of this paper extend the systematic interpretation
of loop diagrams of eleven-dimensional supergravity compactified on $\calT^2$.
The theory obviously has ultraviolet divergences that indicate that the
quantum version of the theory cannot be defined by conventional  quantum field
theory methods.  However, at least for the examples we have studied,
the divergences can be subtracted by introducing cutoff-dependent counterterms
with values that are determined by duality properties expected from the correspondence
with type IIA or IIB string theory on a circle.  Indeed, since the UV divergences are
local they are proportional to the volume of the torus $\calV$ and vanish in the limit
$\calV\to 0$ that corresponds to the ten-dimensional type IIB.

We would now like to analyze the potential divergences at higher orders to see if these
systematics can lead to further insights.  Recall that the eleven-dimensional
one-loop four-graviton amplitude has a superficial UV divergence of order $\Lambda^{11}$
--  where $\Lambda$ is an arbitrary momentum cutoff.  However, the amplitude has an
overall kinematic factor of $\hat K^4$ (where the linearized Weyl tensor, $\hat K$,
 has dimension two) multiplying a scalar field theory box diagram which has a
cubic divergence of order $\Lambda^3$.  Upon compactification there are finite
contributions to this integral  with dependence on the volume determined by
dimensional analysis to be  $1/\calV_n^{3/2}$ (where $n=1$ for a circular
compactification and $n=2$ for $\calT^2$).  These finite terms  come from the
non-zero windings of the loop around homology cycles of $\calT^n$ whereas the
divergent term comes from the zero winding number sector and does not depend on
the moduli of $\calT^n$.
The dilaton-dependent coefficient, $Z_{3/2}$, of the
IIB $\calR^4$ interaction was determined in this way from the one-loop four-graviton
scattering in  \refs{\rfGreenGutperleVanhove}.  At the same time, comparison of the
IIA and  IIB theory at finite $\calV$ resulted in an unambiguous value for the
counterterm that subtracts the $\Lambda^3$ divergence, resulting in a finite value for
the IIA limit, $R_{11}\to 0$.  This fixed the value of the one-loop amplitude of
the IIA theory and hence the $R^4$ interaction in eleven dimensions.

The coefficient, $Z_{5/2}$,  for the
$D^4\calR^4$ interaction was determined in \refs{\gvone} by considering the two-loop
supergravity diagrams, together with a one-loop diagram in which one vertex is
the one-loop counterterm.  The superficial divergence of the
two-loop amplitude is $\Lambda^{20}$.  However, as shown in \refs{\rfBernDunbar},
the two-loop diagrams reduce to a sum of terms in which there is an overall
factor of $S^2 \tilde\calR^4$ (together with the terms with coefficients $T^2
\tilde\calR^4$ and $U^2 \tilde\calR^4$) multiplying a couple of scalar field theory
two-loop diagrams. This external factor has dimension $[\Lambda]^{12}$
and the scalar two-loop diagram  has a new two-loop divergence of order
$\Lambda^8$.  However, it also has
two sub-divergences arising in the sectors where one loop has zero winding.  This
can be subtracted by adding the one-loop diagram in which one vertex is the
one-loop counterterm that is of
order $\Lambda^3$.  This leaves an apparent $\Lambda^5$ divergence.  But there is a
finite term proportional to $\calV_n^{-5/2}$ that arises from non-zero windings of
the other loop around $\calT^n$.
In this case the finite $D^4R^4$ interaction of
the   ten-dimensional type IIB limit ($\calV_2 \to 0$)
arises from a one-loop sub-divergence that is rendered finite by adding the
diagram with the one-loop counterterm.

\subsec{Dimensional analysis and higher order terms}

In this paper we have extended the above analysis to the next order in the momentum
expansion of the two-loop amplitude.  There are now two more external momenta
so the apparent degree of divergence is reduced to $\Lambda^6$.
However, the sectors in
which the loops have non-zero windings around $\calT^2$, which
give a finite contribution proportional to $\calV_n^3$ so in this case we did not
face any divergences -- the integrals were all finite.

We can now ask whether contributions from higher-loop diagrams of
eleven-dimensional supergravity can affect the results we have obtained from
one or two loops.  It was shown in \refs{\rfBernDunbar} that all diagrams beyond one
loop have an external factor of $D^4 \calR^4$.  Using our analysis this means that there
can be no further contributions to the $\calR^4$ term.
This argument motivates the statement
that $\calR^4$  receives no perturbative string
contributions beyond one loop since $Z_{3/2}$
only contains a one string loop contribution \refs{\greengut}.
However, we may well ask whether
the results obtained from two-loop eleven-dimensional supergravity get modified
by eleven-dimensional three-loop and higher-loop effects.  For example,
are higher powers of external momenta (ie, higher numbers of derivatives)  pulled
out into the prefactor multiplying the  sum of diagrams at higher orders?
Unfortunately, the systematics of maximally extended
supergravity is still rather mysterious beyond two loops.
For example, the set of three-loop diagrams
motivated by unitarity cuts in \refs{\rfBernDunbar}  is
incomplete (certain diagrams that have no two-particle cuts are
missing).\foot{MBG is grateful to Lance Dixon for many discussions on this point.}
 It has so far proved too complicated to determine if the complete sum
  has higher powers of momenta in the external prefactor. Such extra external momentum
factors would imply further non-renormalisation theorems.   For example, if the
sum of all three-loop diagrams turned out to have an external factor of $D^8\calR^4$ then it
could not affect the two-loop supergravity calculations of this paper.  This would imply
that the $D^4\calR^4$ and $D^6\calR^4$ interactions would get no further perturbative
contributions beyond two string loops.
The situation with the eleven-dimensional supergravity perturbation theory
would then be analogous to that encountered
in ten-dimensional type II superstring perturbation theory.   There, up
to two loops supersymmetry acts point-wise in the moduli
 space of the world-sheet and
 non-renormalisation statements can then be deduced by knowing the behavior of the
  integrand \refs{\dgp,\dhokerphong,\dhokerphonglecture,\zhou,\BerkovitsMultiloop,\BerkovitsTwoLoop}.  However, for
 three or more loops the integrand contains an explicit overall  factor of $D^4 \, \calR^4$
 \refs{\rfMatone} and any further non-renormalisation theorems, such as for $D^6\, \calR^4$,
  would only be apparent after integrating over the moduli.

However, if we assume that the cut-off procedure outlined
above continues to make sense we can infer some perturbative string theory
non-renormalisation statements
even without detailed knowledge of  the higher-loop terms in supergravity.
For example, we can deduce that the expressions for the
dilaton dependence of the $D^4\calR^4$ and $D^6\calR^4$
interactions presented in \refs{\gvone} and in this paper do not get modified by higher order
terms.
To see this let us focus on the three-loop diagrams.
For simplicity we will consider the case of compactification on a circle of radius
$R_{11}$ to give the IIA theory.
Such diagrams have the superficial degree of divergence of three-loop gravity, which
is $\Lambda^{29}$.  However, we know that the sum of diagrams has a factor
$D^4 \, \calR^4 \sim S^2\calR^4$, which lowers the superficial divergence to $\Lambda^{17}$.
This power can be interpreted by associating $\Lambda^3$ with one one-loop subdivergence, or
$\Lambda^6$ with two one-loop subdivergences,
 or $\Lambda^8$ with the two-loop divergence.  The one-loop
 divergences are regulated by including
 diagrams with the known counterterms\foot{In \refs{\gvone} it was argued that the regulated
 two-loop divergence has to vanish.}.
 Any remaining powers of $\Lambda$ may be transmuted into
inverse powers of $R_{11}$ in the compactified theory.
To make sense in string theory, the result must have an integer power
of the string coupling $g^2_s= R_{11}^3$ in M-theory units (we must also remember the
rule for the Mandelstam invariants, $s=S/R_{11}$, $t=T/R_{11}$ and $u=U/R_{11}$,
 where capital letters
denote the eleven-dimensional invariants).

\ifig\ifthree{The schematic structure of the divergences of three-loop
four-graviton amplitudes compactified
on $\calT^1$ or $\calT^2$.
(a)  The sum of the finite contributions to three-loop diagrams
(only one out of very many
diagrams is pictured).   This should contribute to the $D^{12}\,\calR^4$
interaction.
  (b),(c),(d) The three distinct kinds of two-loop diagrams with a
one-loop counterterm (represented
  by the blob)  that is needed to cancel a one-loop subdivergence. These
contribute to the $D^{10}\, \calR^4$
  interaction.
(e) A one-loop diagram with two one-loop counterterms, which should
contribute to $D^8 \calR^4$.  (f)  A
one-loop diagram with a two-loop counterterm vertex, which was argued to
vanish in \gvone.
(g)  A new primitive three-loop divergence
that makes no contribution in the zero volume limit that gives the
ten-dimensional type  II string theories.
 }{\epsfxsize=10cm \epsfbox{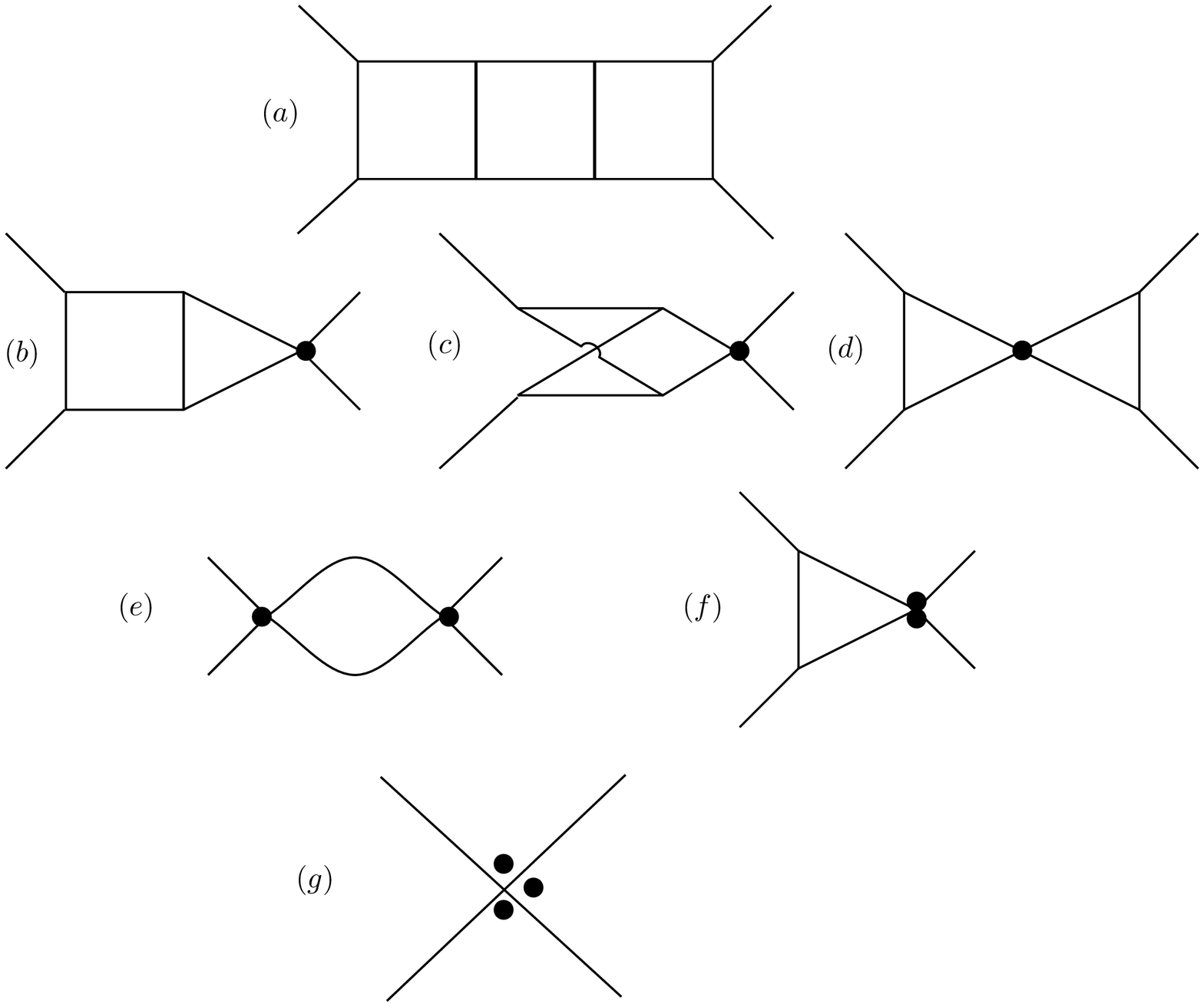}}

As a first example, let us see if there can be a three-loop supergravity
contribution that can be interpreted as a string theory
tree-level $D^4\, \calR^4\sim S^2\, \calR^4$
interaction.  For this to be
the case  we would
need a power of $1/R_{11}^5$ (coming from $1/g^2_s$ and two powers from $s^2=S^2/R^2_{11}$).
We would then reinterpret the 17 powers of $\Lambda$ as $\Lambda^{12}/R_{11}^5$.  However, this would
correspond to four powers of $\Lambda^3$, or four one-loop subdivergences,
 which cannot arise at three
loops!

This argument extends to the $D^6\,\calR^4 \sim S^3\, \calR^4$ interaction.
Now we have two extra powers of external momenta so the superficial degree of divergence
is reduced to $\Lambda^{15}$.  A tree-level contribution would require a power of $1/R_{11}^6$
(taking into account $s^3 = S^3/R_{11}^3$), leaving a net power of $\Lambda^{9}$.
This could only be absorbed by three powers of $\Lambda^3$, but this would require three one-loop
subdivergences.  Such a contribution (see  figure~\ifthree(g))
 would only come from the zero winding number sector
of all three loops and could not produce the requisite dependence on $R_{11}$ that arises
from non-zero windings.

However, there should be a non-zero three-loop supergravity contribution to the $D^8\,\calR^4
\sim S^4\,\calR^4$ interaction, which has a superficial divergence of $\Lambda^{13}$.
In this case a tree-level term behaves as $R_{11}^{-7}$, which leaves a net power of $\Lambda^6$,
which corresponds to two one-loop divergences accompanied by two one-loop counterterms as
shown in~\ifthree(e).
Further analysis of this diagram  strongly suggests that (when compactified on $\calT^2$) it
should contribute the interaction,
\eqn\sevenpo{l_s^6\int d^{10}x\,\sqrt{-g} \, e^{3\phi/2}\, Z_{\sevenh}\, D^8 \calR^4\, ,}
to the effective IIB theory.  The function $Z_{7/2}$ should also follow from supersymmetry.

Likewise,  a contribution of the form
\eqn\tendim{l_s^8\int d^{10}x\, \sqrt{-g}\, e^{2\phi}\, \calE_{(\threeh,\fiveh)}\,
 D^{10}\, \calR^4\, ,}
should arise from the
diagrams containing a counterterm for a single one-loop
subdivergence shown in~\ifthree(b)-(d).  The modular function
$\calE_{(3/2,5/2)}$ is not determined by these very general
arguments, but we know that it
has a tree-level term proportional to $\zeta(3)\zeta(5)$, as can be seen from appendix A.
In principle, this function should again be determined by  supersymmetry using
an extension of the argument of \rfGreenSethi.  In this case there would be an $O(l_s^{16})$
modification of the supersymmetry transformations that would mix $D^{10}\, \calR^4$ with
the Einstein--Hilbert action.  But recall that there are also $O(l_s^6)$ and $O(l_s^{10})$ modifications
to the supersymmetry transformations that mix the
$l_s^{-2}\,Z_{3/2}\, \calR^4$ and $l_s^2\, Z_{5/2}\, D^4\,\calR^4$ interactions (and other interactions
of the same dimension) with the classical action.   These transformations also mix the
$O(l_s^{-2})$ and $O(l_s^2)$ interactions with $l^6\, D^6\, \calR^4$.
It plausibly follows by analogy with our understanding of the
$\calE_{(3/2,3/2)}$ that
$\calE_{(3/2,5/2)}$ satisfies a inhomogeneous Laplace equation of the form
\eqn\threfiv{\Delta_\Omega \calE_{(\threeh,\fiveh)} = A \calE_{(\threeh,\fiveh)} +
B Z_{\threeh}\, Z_{\fiveh}\, ,} where $A$ and $B$ are constants that have not been determined.

Something qualitatively new happens at the next order.
Consider the possible contribution to the $D^{12}\, \calR^4\sim S^6 \, \calR^4$  interaction
coming from compactified eleven-dimensional supergravity.  This interaction
has dimension $[\Lambda]^{20}$,
which reduces the apparent divergence of the three-loop diagrams to $\Lambda^{9}$.
In this case a tree-level term would have a power $1/R_{11}^9$
(noting that $s^6 = S^6/R_{11}^6$) so  the contribution to this interaction
is given by a  {\it finite} integral, which comes from a
sum over all three-loop diagrams (represented by~\ifthree(a)).
However, in this case there are two different
terms in the tree-level expression given in appendix~A.  One of these has coefficient
$\zeta(9)$ and should be associated with  an interaction  of the form
\eqn\ninepo{l_s^{10}\int d^{10}x\,\sqrt{-g} \, e^{5\phi/2}\, Z_{9/2}\, D^{12} \calR^4\, .}
The other tree-level term has coefficient $\zeta(3)^3$ and is associated with an
interaction that could be written as
\eqn\ninepo{l_s^{10}\int d^{10}x\,\sqrt{-g} \, e^{5\phi/2}\, \calE_{(\threeh,\threeh,\threeh)}
\, D^{12} \calR^4\, ,}
where $\calE_{(\threeh,\threeh,\threeh)}(\Omega,\bar\Omega)$ is a new modular form satisfying
a generalized version of our inhomogeneous Laplace equation.

We see, with  the above reasoning, how the structure of the string tree-level four-graviton
amplitude indicates an increasing degeneracy of dilaton-dependent modular functions as the
order in $l_s^2$ increase.  It should be interesting to see how such a structure is in accord
with the constraints of supersymmetry.

\newsec{Summary and discussion of the eleven-dimensional limit}

In this paper we have determined exact properties of the coupling constant dependence of the
$l_s^4\,D^6\, \calR^4$ interaction in the low energy expansion of the type II string theories.
The IIA theory only contains perturbative terms that are proportional to powers of the
string coupling whereas the dependence on the complex coupling in the IIB theory is encoded in
a modular function.  This function,
$\calE_{(3/2,3/2)}(\Omega,\bar\Omega)$, which satisfies a Laplace equation with a
source term \eLaplaceEq\
is given in \eIfinala\ as the sum of two terms, $S$ an $R$.
The first part $S$ is an infinite series of terms proportional to
$\calZ_{(p+3/2,p+3/2)}$ ($p=0,1,\dots$) defined by \eCalZ\ and $R$ is given by \rdeff.

We discussed properties of the function $\calE_{(3/2,3/2)}$ in terms of its Fourier
modes that are proportional to $e^{2\pi i k\Omega_1}$.  This is the decomposition into sectors of different
$D$-instanton charges, $k$.
The zero mode sector contains the `perturbative' terms that are powers of $\Omega_2\equiv
e^{-\phi}$.  We determined the coefficients of the four perturbative terms,
which are proportional
to $\Omega_2^3$, $\Omega_2$, $\Omega_{2}^{-1}$ and $\Omega_2^{-3}$,
corresponding to tree-level, one-loop, two-loop and three-loop
contributions in string perturbation theory, respectively.
The tree-level and one-loop terms precisely match string
perturbation theory results whereas the
value of the two-loop coefficient (the term with $48/5\, \zeta(2)^2$ in~\eIonepert)
and the three-loop coefficient (the value of $\beta$ given in \eBeta)
 have not yet been calculated in string theory and so are predictions.  These properties
followed by analyzing the Laplace equation rather than its explicit
solution,  together with the boundary condition that the leading power of $\Omega_2$
contained in $\calE_{(3/2,3/2)}$
is the tree-level term in the weak coupling limit (which is easy to verify from
the explicit solution).  For completeness, we also extracted the three-loop
coefficient of the type IIA theory from a formula in \refs{\gvone}  and found that this was
also equal to $\beta$.  This may be of interest since
equality of the four-graviton amplitudes in the IIA and IIB
theories is not obvious beyond two loops

The non zero-mode sector includes
an infinite series of exponentially suppressed $D$-instanton terms
that are proportional to $e^{2\pi i k \Omega}$, where $k\ne 0$ is the $D$-instanton charge.  There is
also an infinite series of terms corresponding to pairs of $D$-instantons with charges $k_1$ and $k_2$.
Each term is proportional to  $e^{2\pi i (k_1+k_2) \Omega_1}\, e^{-2\pi i (|k_1|+|k_2|)\Omega_2}$.
In general for these terms the magnitude of the $D$-instanton charge is smaller than the action since
$|k_1+ k_2| \le |k_1|+|k_2|$.  Equality only holds for the cases where both $k_1$ and $k_2$ have
the same sign.  The terms in which $k_1 = -k_2 \equiv \hat k$ again
contribute to the zero mode of $\calE_{(3/2,3/2)}$, this time with  exponentially decreasing
factors proportional to $e^{-4\pi \hat k \Omega_2}$.

The above features were obtained by extending previous work on the duality between the
four-graviton scattering amplitude of eleven-dimensional supergravity compactified on $\calT^2$
with the amplitude in type IIB string theory.
In particular, we extended the work of \gvone\ which deduced the
dilaton-dependent prefactor, $Z_{5/2}$ of the $l_s^2 \, D^4\, \calR^4$ interaction from the
low energy limit of two-loop effects in eleven dimensions.  In this paper we extracted the next term in
the momentum expansion of two-loop eleven-dimensional supergravity on a circle (to get
to IIA) as well as on $\calT^2$ (to get to IIB).
In both cases the string theory terms we are interested in emerge in the zero volume limit ($R_{11}\to
0$ for IIA and $\calV\to 0$ for IIB).  Although the tree-level IIA string
amplitude is obtained by this procedure, the loop corrections arise from
undetermined divergent contributions.
In contrast,  the full modular function of the IIB theory is determined entirely by finite integrals
(whereas in \gvone\ we needed to consider a particular subdivergence).   However,
it is well known that the
four-graviton amplitudes in the IIA and IIB string theories have the same perturbative expansion,
at least up to two loops (beyond that there is the possibility of contributions from
odd-odd spin structures that have opposite signs in the two theories).  Therefore,
the perturbative terms in the $D^6\, \calR^4$ interaction of the IIA theory are determined
once they are given in the IIB theory.

We also argued  that the structure of the Laplace equation is intuitively that
expected from a generalization of the supersymmetry arguments in \rfGreenSethi, which determined the
coefficient of $\calR^4$.   The source term  in the Laplace equation,
$Z_{3/2}^2$ arises from the fact that terms in
effective action of the same order as
 $l_s^2\, D^6\,\calR^4$ are not only related by supersymmetry to the classical action but also to
 $l_s^{-2}\, \calR^4$ and associated terms of the same dimension.  It would be good to
 make this argument more precise.  Following this line of reasoning, at the next order in $l_s$
supersymmetry cannot mix the  $l_s^6\, D^8\, \calR^4$ with anything and its prefactor
should satisfy a homogeneous Laplace equation with solution $Z_{7/2}$.
However, as argued in section 6,  at the next order another inhomogeneous
Laplace equation,  with source $Z_{3/2}\, Z_{5/2}$~\threfiv,
 should determine the dilaton-dependent
 prefactor $\calE_{(3/2,5/2)}$ of the $l_s^8\, D^{10}\,\calR^4$ interaction.
The order after that reveals new systematics.  This is the first time that two distinct
terms in the expansion of the tree-level amplitude contribute -- one with coefficient $\zeta(9)$
and the other $\zeta(3)^3$.  This presumably indicates a branching, with two distinct  modular
functions arising in the prefactor.  It is clearly of interest to study systematics of the exact
four-graviton amplitude at this order and beyond.

Finally, let us comment on the  eleven-dimensional limit.
In the case of the $\calR^4$ interaction the eleven-dimensional limit was determined entirely
in terms of the coefficient of the one-loop amplitude \refs{\rfGreenVanhoveMtheory}.
  Similarly the value of the eleven-dimensional limit of the $D^6\,\calR^4$ interaction
  is determined by the
two-loop contribution  in the IIA theory.  Since this is the same as the two-loop
contribution to the IIB theory it is given by the $\Omega_2^{-1}$ term in $\calE_{(3/2,3/2)}$.
Making use of the dictionary in appendix~B and the fact that the IIA two-loop term is of order
$R_{11}^3$ results in the contribution to the eleven-dimensional action,
\eqn\mthres{S =l_{11}^{3}\, {\zeta(2)^2\over120\cdot (4\pi)^7}\,
 \int d^{11}x\, \sqrt{-G}\, D^6\, \calR^4\, .}
The fact that the
 $D^6\,\calR^4$ interaction has a finite eleven-dimensional limit whereas the $D^4\calR^4$
 interaction is absent in eleven dimensions is in accord with analogous statements
  concerning powers of the curvature.  The first power of $\calR$ after $\calR^4$ that contributes
in eleven  dimensions was conjectured   in
\refs{\rfRussoTseytlin} to be $\calR^7$,
which has the same dimension as $D^6\calR^4$.   This seems also to be in
accord with a very mysterious observation of \refs{\DamourZB}
based on representations of $E_{10}$.

\newsec{Acknowledgements}
We are grateful to Savdeep Sethi for useful insights into the role
 of supersymmetry
and to Lance Dixon and David Kosower for conversations on higher-loop supergravity.
P.V. thanks the LPTHE of Jussieu for hospitality when this work was done.
P.V. was partially supported by EU Research Training Networks MRTN-CT-2004-503369 and  MRTN-CT-2004-005104 and MBG by MRTN-CT-2004-512194.


\appendix{A}{Review of tree-level string amplitude and expansion of $Z_s(\Omega,\bar\Omega)$}

{\it  (i) Tree-level four graviton scattering in type II string theory}

The amplitude has the form
\refs{\rfGreenSchwarz,\rfGreenSchwarzWitten},
\eqn\etreeone{
A^{(2)}_4 =\kappa_{10}^2\,  \hK \, e^{-2\phi}\, T(s,t,u)\, }
where $2\kappa_{10}^2= (2\pi)^7  l_s^8 $ and $T(s,t,u)$
 is given by
\eqn\eTree{
\eqalign{
T &= {64\over l_s^6  stu}
{\Gamma(1-{l_s^2\over4} s)\Gamma(1-{l_s^2\over 4} t)\Gamma(1-{l_s^2\over 4} u)
 \over \Gamma(1+ {l_s^2\over 4} s)\Gamma(1 + {l_s^2\over 4} t)\Gamma(1  +
 {l_s^2\over 4} u)} \cr
&= {64 \over l_s^6 stu} \exp\left(\sum_{n=1}^\infty {2 \zeta(2n+1) \over
 2n+1}{l_s^{4n+2}\over 4^{2n+1}} (s^{2n+1} + t^{2n+1} + u^{2n+1})\right)\ .
}}
The low energy expansion of the amplitude begins with the
terms, (making use of some identities proved in section~2 of \refs{\gvtwo})
\eqn\etreeexp{
\eqalign{  T & = {64\over l_s^6 stu}   +
2\zeta(3)   +  {\zeta(5)\over16}l_s^4\, (s^2+t^2+u^2)\cr
& + {\zeta(3)^2\over 96}  l_s^6\, (s^3+t^3+u^3) +{\zeta(7)\over 512}
 l_s^8 (s^2+t^2+u^2)^2  + {\zeta(3)\zeta(5) \over 1280}
   l_s^{10} \,(s^5+t^5+u^5)+ \cr
   & + {\zeta(9)\over 4096}\, l_s^{12}\, \left({2\over 81}(s^6+t^6+u^6) +
    {7 \over 108}(s^2+t^2+u^2)^3\right)\cr
   & + {\zeta(3)^3\over 4096}\, l_s^{12}\,
{4 \over
27} (s^2+t^2+u^2)^3
 + \dots \ .
}}
For our considerations it is notable that the $l_s^6\,\zeta(3)^2 (s^3+t^3+u^3)$ term is the
first that is not linear in the exponent of \eTree.  Also, note that the first degeneracy
of terms at a given dimension arises at order  $l_s^{12}$, where there is a contribution with
coefficient $\zeta(9)$ and one with coefficient $\zeta(3)^3$.

In terms of the coordinates of the
eleven-dimensional theory \refs{\gvone} the  expression \etreeexp\  has the
low-energy expansion,
\eqn\emexp{ \eqalign{{T \over R_{11}^{3}}  & =  {64\over l_{11}^6 STU} +
{2\zeta(3)\over R_{11}^3}   + {\zeta(5)\over 16} {l_{11}^4\over R_{11}^5}
(S^2+T^2+U^2)   \cr
&+  {\zeta(3)^2\over 96 }
{l_{11}^6\over R_{11}^6}\, (S^3+T^3+U^3)+{\zeta(7)\over 512 }{l_{11}^8\over R_{11}^7}\, (S^2 + T^2 + U^2)^2 +
 {\zeta(3)\zeta(5) \over 1280}{ l_{11}^{10}\over  R_{11}^{8} }, (S^5+T^5+U^5)\cr
& + {\zeta(9)\over 4096}\, {l_{11}^{12}\over R_{11}^{9}}
\left({2 \over 81}\,(S^6+T^6+U^6) +{7\over 108}\, (S^2+T^2+U^2)^3\right)\cr
& + {\zeta(3)^3\over 4096}\, {l_{11}^{12}\over R_{11}^{9}}\left({4\over
27}\, (S^2+T^2+U^2)^3\right)
\dots
\,.}}

\medskip

{\it (ii) Expansion of the non-holomorphic Eisenstein series  $Z_s(\Omega,\bar\Omega)$}

 For general values of $s$ the expression \eisendef\ can be
expanded as a Fourier series,
\eqn\asymfour{
 Z_s (\Omega,\bar \Omega)   = \sum_k \calZ^s_k e^{2\pi i k \Omega_1},}
 where
 \eqn\asymzero{ \calZ^s_0  = 2\zeta(2s)
\Omega_2^{s}   + 2\sqrt\pi\,  \Omega_2^{1  -s} {
\Gamma( s-\half)\zeta(2s-1) \over \Gamma(s)}}
  and
\eqn\asymk{\calZ^s_k
= {4\pi^{s}\over \Gamma(s)} \, |k|^{s-\half}
\,\mu(k,s)\, \Omega_2^{{1\over 2}} \calK_{s-\half} (2\pi|k|\Omega_2)\,
e^{2\pi i k\Omega_1},}
with
\eqn\measuredef{\mu(k,s) =\sum_{\hat m|k} \hat m^{-2s+1}.}
 The  modified Bessel
function ${\cal K}_s$  has the
 integral representation
\eqn\bessdef{ \calK_s(z) = {1\over 2} \left( {z\over 2}\right)^s
\int_0^\infty {dt \over t}t^{-s}e^{-t -z^2/4t}}
  and the
asymptotic expansion for large $z$,
\eqn\asymdef{\eqalign{\calK_s
(z) =& \left(\pi\over 2z\right)^{1/2} e^{-z}  \sum_{k=0}^\infty
{1\over (2z)^k }{\Gamma(s + k + {1\over 2}) \over \Gamma(k+1)
\Gamma(s - k + {1\over 2}) }\ .  \cr}}
Substituting this expansion in
\asymfour--\asymk\  leads to the series \asymterm.

\appendix{B}{The dictionary between  supergravity and superstring theories}

In order to compare the results obtained from compactified
eleven-dimensional supergravity with those of the IIA or IIB string
theories we here review the dictionary that relates the parameters in the
various descriptions.
Compactification on a circle of radius $R_{11}$
 gives rise to the type IIA string theory
where  the string coupling constant,
$g^A=e^{\phi^A}$ (where $\phi^A$ is the IIA dilaton),
is given  by $l_{11} = (g^A)^{1/3} \ls$ and
$R_{11}^3=e^{2\phi^A} =(g^A)^2$. Masses are measured with the metric \refs{\rfWittenVarious}
\eqn\eMetric{
ds^2=G^{(11)}_{MN}dx^Mdx^N=
 {l_{11}^2\over l_s^2R_{11}}\; g_{\mu \nu} dx^\mu dx^\nu +
R_{11}^2 l_{11}^2 (dx^{11} - C_\mu dx^\mu )^2,
}
where $g_{\mu \nu}$ is the string frame metric.
Since the compactification radius
$R_{11}$ depends on the string coupling constant the Kaluza-Klein modes are
mapped to the massless fundamental string states  and
the non-perturbative D0-brane states.
When expressed in terms of the type IIA string theory parameters the
compactified classical action becomes
\eqn\iiaclass{S_{EH} = {1\over 2 \kappa_{10}^2} \int d^{10} x\, \sqrt{- g} \,
e^{-2\phi^A}\, R,}
where  $2\kappa_{10}^2=(2\pi)^7 l_s^8$ and $l_s$ is the string length
scale.\foot{In this convention the fundamental string tension is
related to the string scale by
$T_F^2 = \pi (2\pi l_s)^4/\kappa_{10}^2$.}

More generally, we want to
consider compactification of the eleven-dimensional theory on a two-torus of volume
$\calV$ and complex structure $\Omega$ and compare with type IIA string theory
compactified on a circle of radius $r_A$  and type IIB compactified on a circle of
radius $r_B = 1/r_A$ (where these radii are dimensionless quantities that are defined in the
respective string frames).
 The dictionary
that relates $\calV$ and $\Omega$ to the nine-dimensional type IIA and type IIB
string theory parameters is  \refs{\rfAspinwall,\rfSchwarzPower},
\eqn\usdico{\eqalign{
\calV=  &R_{10}R_{11}= \exp\left({1\over 3}\phi^B\right)
r_B^{-{4\over 3}},\qquad   r_B={1\over R_{10} \sqrt{R_{11}}} =
r_A^{-1},
\cr
&  \Omega_1 = C^{(0)} = C^{(1)}_9, \qquad  \Omega_2= {R_{10}\over
 R_{11}}=\exp\left(-\phi^B\right)=r_A\, \exp\left(-\phi^A\right).\cr}}
The one-form $C^{(1)}$ and the zero-form $C^{(0)}$
are the respective \RR\ potentials and $\phi^A$, $\phi^B$ are the
IIA and IIB dilatons.

In the text we use this dictionary to convert the leading contribution to the effective $D^6\, \calR^4$
M-theory action in the limit $\calV\to 0$, which behaves as $\calV^{-3}$, to the corresponding
action of ten-dimensional type IIB string theory, which  is a finite quantity
in the $\calV\to 0$ limit.

\appendix{C}{Laplace Equations }

\subsec{Laplace equation for $A(\tau,\bar\tau)$}

In this section we derive the Laplace equation~\lapa\ satisfied by
$A(\tau)=\hat A(|\tau_{1}|+i\tau_{2})$ defined in~\adef.
We consider the following integral over the fundamental domain for $Sl(2,\ZZ)$
\eqn\eIntegra{
I= \int_{\calF}\, {d^2\tau\over\tau_{2}^2}\, A(\tau)\, \Delta_{\tau}
F(\tau,\bar\tau)
}
where $F(\tau,\bar\tau)$ is an arbitrary modular invariant function,
 exponentially decreasing for $\tau_{2}\to\infty$
 (which is the case for $\exp(-\pi E)$ in~\newexp\ for non vanishing
 $m$ and $n$). Integrating by part one should pay attention to the
 fact that because
of the absolute value on $\tau_{1}$ there are boundary contributions
 from $\tau_{1}=0$. Therefore one gets,
\eqn\eIntII{\eqalign{
I& = \int_{\calF}\, {d^2\tau\over\tau_{2}^2}\, \Delta_{\tau}A(\tau)\,
 F(\tau,\bar\tau)\cr
&+ \int_{\partial_{\tau_{1}}{\cal F}} d^2\tau \,
\left[-\partial_{\tau_{1}}A(\tau) \,F(\tau,\bar\tau)+
 A(\tau)\partial_{\tau_{1}}\,F(\tau,\bar\tau)\right] \cr
&+ \int_{\partial_{\tau_{2}}{\cal F}} d^2\tau \,
\left[-\partial_{\tau_{2}}A(\tau) \,F(\tau,\bar\tau)+
 A(\tau)\partial_{\tau_{2}}\,F(\tau,\bar\tau)\right] \ .
}}
By modular invariance and the fact that $F(\tau)$ is exponentially decreasing for $\tau_{2}\to\infty$,
 the $\tau_{2}$-boundary $\partial_{\tau_{2}}\calF$ does not contribute. From the
  $\tau_{1}$-boundary, the modular properties of $A$ assures that only  the boundary $\tau_{1}=0$ contributes, and
\eqn\eIbndry{\eqalign{
\delta I &= -12 \,  \int_{1}^\infty  d\tau_{2} \, \delta(\tau_{1})\, \left(-{1\over \tau_{2}} \right) \, F(\tau,\bar\tau)\cr
&=-12   \int_{\calF}  {d^2\tau_{2}\over\tau_{2}^2}\, \tau_{2}\delta(\tau_{1})\, F(\tau,\bar\tau)\ .
}}
In the interior of the fundamental domain where $\tau_{1}\neq 0$ one easily derives that $\Delta_{\tau}A(\tau)=\Delta_{\tau}\hat A(\tau)=12\, A(\tau) $
Therefore the integral $I$ is given by
\eqn\eIFinal{
I = \int_{\calF}\, {d^2\tau\over\tau_{2}^2}\,
(12\, A- 12 \tau_{2}\delta(\tau_{1}))\, F(\tau,\bar\tau)
}

\subsec{Laplace equation for $\calZ_{(s,s)}$}

Equation \eZZDelta\ is obtained by first noting the following identities,
\eqn\eZeZ{\eqalign{
\Omega_{2}\partial_{\Omega} Z_{s}^{(\hm_{I})}&={s\over 2i}\,
 {\Omega_{2}^{s}\over |\hm_1+\hm_2\Omega|^{2s}}\, {\hm_1+\hm_2\bar\Omega\over \hm_1+\hm_2\Omega}\cr
\Omega_{2}\bar\partial_{\bar\Omega} Z_{s}^{(\hm_{I})}&=- {s\over 2i}\,
{\Omega_{2}^{s}\over |\hm_1+\hm_2\Omega|^{2s}}\, {\hm_1+\hm_2 \Omega\over \hm_1+\hm_2\bar\Omega}\, ,}}
and
\eqn\eZZz{
(\hm_1+\hm_2\Omega)^2 (\hn_1+\hn_2\bar\Omega)^2 + c.c.  = 2|\hm_1+\hn_2\Omega|^2 \,
 |\hn_1+\hn_2\Omega|^2 -4 (\hm_1\hn_2-\hn_1\hm_2)^2\, \Omega_{2}^2
\, ,}
where the quantity $Z_{s}^{(\hm_{I})}$ is defined in \eZZ.
It follows that
\eqn\ZZee{
\Delta_{\Omega}   ( Z_{s}^{(\hm_{I})} \, Z_{s'}^{(\hn_I)})= (s+s')(s+s'-1)\,Z^{(\hm_{I})}_{s}
Z^{(\hn_{I})}_{s'} - 4ss'\, (\det M)^2 \, Z^{(\hm_{I})}_{s+1}\, Z^{(\hn_{I})}_{s'+1}\, ,}
so that, after multiplying by $(\det M)^{s+s'}$ and
 summing over the integers $\hm_1$, $\hm_2$, $\hn_1$ and $\hn_2$,
 the generalized series  $\calZ_{(s,s')}$ defined in \eCalZ\ is found to satisfy the differential equation
 \eZZDelta,
\eqn\eEEnn{
\Delta_{\Omega}\calZ_{(s,s')} = (s+s')(s+s'-1)\, \calZ_{(s,s')} - 4ss' \, \calZ_{(s+1,s'+1)}\, .
}
\subsec{Laplace equation for the lattice sum}

We here produce some details of the calculations used in the main body of the text.
The exponent has the expansion
\eqn\eewert{\eqalign{
E & = {\calV V\over \Omega_2 \tau_2} \left|(1\ \Omega)\, M \, (\tau\
  1)\right|^2 + 2\calV V \det M \cr
  &=  {\calV V \over \Omega_2 \tau_2} \, \left[|m_1 + m_2\Omega|^2 + |\tau|^2 |n_1 + n_2 \Omega|^2
+ 2\tau_1 ((n_1 + n_2 \Omega_1)(m_1 + m_2 \Omega_1) + n_2m_2 \Omega_2^2)\right]\,
}}

We will write
\eqn\xndef{ E = {X\over \tau_2}\,,}
so that
\eqn\propsx{\eqalign{\partial_{\tau_1}^2 E &={\partial_{\tau_1}^2 X\over \tau_2}\,
\qquad
\partial_{\tau_2}E  = -{E\over \tau_2}+ {1\over \tau_2}\partial_{\tau_2}X \, , \qquad
\partial_{\tau_2}^2 E = {2\over \tau_2^2}\left(E-  \partial_{\tau_2}X\right)\, ,}}
where
\eqn\xress{\partial_{\tau_2} X = {2\calV V\over \tau_2} \,
\tau_2 |n_1 + n_2\Omega|^2\, .}
As a result we have
\eqn\moredefs{\eqalign{
\tau_2^2\, \partial_\tau^2 E &= 2E - 2\partial_{\tau_2}X + \tau_2^2\partial_{\tau_1}^2
X = 2E\, ,
\cr
\qquad \tau_2^2\,\partial_{\tau }E \cdot \partial_{\tau }E &=
   E^2 + (\partial_{\tau_1}X)^2+  (\partial_{\tau_2 }X)^2 - 2E  \partial_{\tau_2 }X
= E^2 - 4\calV^2 V^2\, (\det M)^2 \, ,}}
where the explicit form of $E$ has been used on the right-hand side of these equations,
Therefore, we have that
\eqn\totlap{\Delta_{\Omega} e^{-\pi E} \equiv 4 \Omega_2^2\, \partial_\Omega\,
  \partial_{\bar \Omega}\, e^{-\pi E} = \pi E^2 -2\pi E  -(2\calV V\,\det M)^2 \, ,}
as quoted in the text.

 \listrefs
\bye